\documentclass[12 pt]{article}
\usepackage{times,bm}
\usepackage[affil-it,]{authblk}
\usepackage[pagebackref=false,colorlinks=true,linkcolor=blue,citecolor=blue,urlcolor=blue]{hyperref}
\usepackage{tensor}
\usepackage{geometry}
\usepackage{graphicx}
\usepackage{cite}
\usepackage{romannum}
\usepackage{amsmath}
\usepackage{breqn}
\usepackage{amsfonts}
\usepackage{soul}
\usepackage{amssymb}
\usepackage{subfigure}
\usepackage{color}
\usepackage{float}
\usepackage{multirow,multicol}
\usepackage{tabulary}
\usepackage[flushleft]{threeparttable}
\usepackage{pgfplots,subfigure}
\usepackage{xcolor}
\numberwithin{equation}{section}
\usepackage{tikz}
\usepackage{placeins}
\usepackage{caption}
\usepackage[normalem]{ulem}
\usepackage{hyperref}
\usepackage{enumitem}
\usepackage{nameref}
\usepackage{comment}
\usepackage{pgfplots}
\pgfplotsset{compat=1.7}

\usepgflibrary{arrows}
\usetikzlibrary{shapes.callouts}
\tikzset{
	level/.style   = { thick, },
	connect/.style = { dotted, red   },
	notice/.style  = { draw, rectangle callout, callout relative pointer={#1} },
	label/.style   = { text width=2cm }
}
\newcommand{\ie}{\begin{equation}}
\newcommand{\fe}{\end{equation}}

\usepackage{letltxmacro}
\makeatletter
\let\oldr@@t\r@@t
\def\r@@t#1#2{%
	\setbox0=\hbox{$\oldr@@t#1{#2\,}$}\dimen0=\ht0
	\advance\dimen0-0.2\ht0
	\setbox2=\hbox{\vrule height\ht0 depth -\dimen0}%
	{\box0\lower0.4pt\box2}}
\LetLtxMacro{\oldsqrt}{\sqrt}
\renewcommand*{\sqrt}[2][\ ]{\oldsqrt[#1]{#2}}
\makeatother

\begin{document}
\pagenumbering{arabic}

	\newcommand{{\ri}}{{\rm{i}}}
	\newcommand{{\Psibar}}{{\bar{\Psi}}}
	\newcommand{{\red}}{\color{red}}
	\newcommand{{\blue}}{\color{blue}}
	\newcommand{{\green}}{\color{green}}
	\newcommand{\rev}[1]{\textbf{\textcolor{red}{#1}}}
	
	\title{Absorption, Scattering, Geodesics, Shadows and Lensing Phenomena of Black Holes in Effective Quantum Gravity}

	\author{\large  
		\textit {N. Heidari}$^{\ 1}$ \footnote{E-mail: heidari.n@gmail.com (Corresponding author)},
        \textit{A. A. Ara\'{u}jo Filho}$^{\ 2}$
		\footnote{E-mail: dilto@fisica.ufc.br},  
		\textit {R. C. Pantig}$^{\ 3}$\footnote{E-mail: rcpantig@mapua.edu.ph},
         \,
		\textit {A. Övgün}$^{\ 4}$\footnote{E-mail: ali.ovgun@emu.edu.tr},
		\\
		\small \textit {$^{\ 1}$Center for Theoretical Physics, Khazar University, 41 Mehseti Street, Baku, AZ-1096, Azerbaijan.}\\
        \small \textit {$^{\ 2}$Departamento de Física, Universidade Federal da Paraíba, Caixa Postal 5008, 58051-970, João Pessoa, Paraíba,  Brazil.}\\
  
		\small \textit {$^{\ 3}$Physics Department, Map\'ua University, 658 Muralla St., Intramuros, Manila 1002, Philippines.}\\

        \small\textit {$^{\ 4}$ Physics Department, Eastern Mediterranean University,
        Famagusta, 99628 North Cyprus, via Mersin 10, Turkiye.}
	}

	\date{\today}
	\maketitle

\begin{abstract}

In this work, we investigate the signatures of black holes within an effective quantum gravity framework recently proposed in the literature \cite{zhang2024black}. We begin by outlining the general setup, highlighting the two distinct models under consideration. This includes a discussion of their general properties, interpretations, and the structure of the event and inner horizons. We then examine the behavior of light in this context, analyzing geodesics, the photon sphere, and shadow formation. To validate our results, we estimate lower bounds for the shadow radius based on observational data from Sgr A$^{*}$ and M87$^{*}$. Subsequently, we derive the partial radial wave equation for scalar perturbations, enabling us to study the absorption cross--section in both low-- and high--frequency regimes. Additionally, we evaluate the greybody factors and provide bounds for both bosonic and fermionic fields. Finally, we present a detailed analysis of gravitational lensing in both the weak and strong deflection limits. For the weak deflection regime, the \textit{Gauss--Bonnet} theorem is employed, while for the strong deflection limit, the \textit{Tsukamoto} approach is utilized.

\end{abstract}
%%%%%%%%%%%%%%%%%%%%%%%%%%%%%%%%%%%%%%%%%%%%%%%%%%%%%%%%%%%%%%%%%%%%%%%%%%%%%%%%%%%%%%%%%%%%%%%%%%%%%%%%%%%%%%%%%%%%%%%%%%%%%%%%%%%%%%%%%%%%%%%%%%%%%%%%%%%%%%%%%%%%%%%%%%%%%%%%%%%%%%%%%%%%%%%%%%%%%%%%%%%%%%%%%%%%%%%%%%%%%%%%%%%%%%%%%%%%%%%%%%%%%%
	\begin{small}
	\end{small}
	\tableofcontents
	\FloatBarrier
	
%%%%%%%%%%%%%%%%%%%%%%%%%%%%%%%%%%%%%%%%%%%%%%%%%%%%%%%%%%%%%%%%%%%%%%%%%%%%%%%%%%%%%%%%%%%%%%%%%%%%%%%%%%%%%%%%%%%%%%%%%%%%%%%

\section{Introduction}

General relativity (GR), despite its remarkable achievements, encounters several obstacles, such as the emergence of singularities \cite{ll1} and its incompatibility with quantum theory \cite{ll2,ll3,ll4,ll5}. These limitations indicate that GR may not represent the definitive description of spacetime. Consequently, ongoing research aims to extend or modify GR, often by altering one of the fundamental assumptions outlined in the Lovelock theorem, which establishes the uniqueness of Einstein’s formulation of GR \cite{ll6}. In addition to various modifications discussed in the literature \cite{ll7,ll8,ll9,battista2024quantum,donoghue2023quantum}, this letter investigates a distinct approach that leverages the Hamiltonian formalism to construct an alternative to GR while preserving the 4--dimensional diffeomorphism invariance. Such a strategy is particularly applicable when a canonical quantum gravity framework results in a semiclassical gravity theory derived from its Hamiltonian formulation.

In GR, diffeomorphism covariance translates into the structure of the Poisson algebra when the theory is recast in its Hamiltonian formulation. However, a reverse challenge arises: determining under what circumstances a given $3 + 1$ Hamiltonian model corresponds to a generally covariant spacetime theory. This question, first explored in works such as \cite{ll10}, is a fundamental concern for any effective Hamiltonian framework that emerges from a canonical approach to quantum gravity. For example, this problem has been extensively debated in the context of effective models derived from symmetry--reduced versions of loop quantum gravity \cite{ll11,ll12,ll13,ll14,ll15,ll16,ll17,ll18,ll19,ll20,ll21,ll22,ll23,ll24,ll25,ll26,ll27,ll28,ll29,ll30,ll31,ll32,ll33,ll34,ll35,ll36,ll37,ll38}.

Previous studies \cite{ll16,ll20,ll27,ll29,ll31,ll33,ll39} have typically addressed the covariance problem by selecting a particular matter field to fix the diffeomorphism gauge. In these cases, quantization produces effective Hamiltonians that are applicable only within the chosen gauge. In contrast, a recent approach in the literature \cite{zhang2024black} has taken steps to move away from this reliance on matter field gauge fixing and aims to construct effective Hamiltonian constraints that remain consistent across different gauges. This work specifically focused on spherically symmetric vacuum gravity, establishing the conditions for maintaining general covariance, finding the solutions that satisfy these conditions, and examining the corresponding spacetime metrics that emerge.

The field of cosmological research has greatly advanced, thanks in part to the recent detection of gravitational waves by experiments such as the LIGO--Virgo collaborations \cite{018,016,017}. Gravitational waves have become indispensable for probing various cosmic phenomena, including the analysis of gravitational lensing effects within the weak--field approximation \cite{heidari2023gravitational,020,araujo2024gravitational,019,aa2024implications,araujo2023analysis}. Historically, studies on gravitational lensing have primarily focused on light propagation over extensive distances from gravitational sources, starting with the Schwarzschild geometry \cite{021} and later extending to more general spherically symmetric and static spacetimes \cite{022}. However, in regions characterized by intense gravitational fields, such as those surrounding black holes, the deflection angle of light is greatly amplified, as expected in strong--field scenarios. Moreover, the deflection of light also leads to the formation of the black hole shadow, which represents one of the most striking phenomena in astrophysics, offering a direct glimpse into the warped fabric of spacetime near a black hole's event horizon. This shadow is essentially a dark silhouette against the glowing background of hot accreting material, first theoretically predicted by James Bardeen in the 1970s as part of the Kerr metric analysis \cite{Cunningham}. It gained prominence with the foundational work of Falcke, Melia, and Agol in 2000 \cite{Falcke:1999pj}, who proposed imaging the shadow of the supermassive black hole at the center of our galaxy, Sagittarius A*, using Very Long Baseline Interferometry (VLBI). The observational breakthrough came in 2019 when the Event Horizon Telescope (EHT) collaboration presented the first image of a black hole shadow in the galaxy M87 and Sgr. A*, validating key aspects of General Relativity and opening a new era in studying black hole physics, accretion processes, and testing gravity theories in extreme conditions \cite{Vagnozzi:2019apd, Bambi:2019tjh, Allahyari:2019jqz, Kumar:2020hgm, Afrin:2021imp, Khodadi:2021gbc, Afrin:2021wlj, Khodadi:2022pqh, Fu:2021fxn, Afrin:2022ztr, Afrin:2023uzo, Ghosh:2022kit, Afrin:2024khy, Khodadi:2024ubi}.

The groundbreaking observations of the supermassive black hole at the center of the M87 galaxy, captured by the Event Horizon Telescope, have opened new avenues for research and drawn significant attention from the scientific community \cite{EventHorizonTelescope:2019ggy,028,025,024,027,026}. A key development in this area was introduced by Virbhadra and Ellis, who formulated a simplified lens equation suited for studying supermassive black holes in asymptotically flat spacetimes \cite{031,virbhadra2000schwarzschild}. Their findings revealed that strong gravitational lensing generates multiple symmetric images around the optical axis. Building on this, Fritelli et al. \cite{032}, Bozza et al. \cite{033}, and Tsukamoto \cite{035} further refined the analytical tools for investigating strong--field lensing phenomena. A wide range of configurations have since been explored, including light bending in various gravitational backgrounds \cite{Virbhadra:2022ybp,virbhadra2000schwarzschild,grespan2023strong,cunha2018shadows,Kuang:2022xjp,metcalf2019strong,Ovgun:2018tua,virbhadra2002gravitational,bisnovatyi2017gravitational,aa2024antisymmetric,ezquiaga2021phase,virbhadra1998role,Okyay:2021nnh,Li:2020dln,Ovgun:2018fnk,Pantig:2022gih,Pantig:2022ely,oguri2019strong}, alternative gravity theories \cite{nascimento2024gravitational,heidari2023gravitational,chakraborty2017strong,40}, exotic structures such as wormholes \cite{38.5,ovgun2019exact,38.2,38.1,38.4,38.3}, charged solutions like the Reissner--Nordström black hole \cite{036,036.2,036.1} and rotating spacetimes \cite{jusufi2018gravitational,hsieh2021strong,37.1,37.2,hsieh2021gravitational,37.4,37.3,37.6,37.5}. Additionally, research on gravitational distortions and the related optical phenomena continues to expand this area of study \cite{virbhadra2022distortions,virbhadra2024conservation}.

Following the solution presented in \cite{zhang2024black}, several recent studies have emerged in the literature exploring various properties of the model, including gravitational lensing in both the weak \cite{Li:2024afr} and strong \cite{liu2024gravitational} deflection regimes using Bozza’s method \cite{bozza2002gravitational}, light rings \cite{liu2024light}, quasinormal modes of the Dirac field \cite{malik2024perturbations}, correlations between quasinormal modes and shadows \cite{konoplya2024probing}, long-lived modes \cite{Bolokhov:2024bke}, and the Hod’s bound \cite{Malik:2024elk}. However, up to date, a comprehensive analysis that addresses geodesic trajectories, a detailed analytical study of the photon sphere and shadow structures, absorption cross--sections,  scattering properties, greybody factor for scalar and Dirac perturbation, and gravitational lensing for both weak (using the \textit{Gauss--Bonnet} theorem \cite{gibbons2008applications}) and strong deflection limits (through \textit{Tsukamoto’s} technique \cite{tsukamoto2017deflection}) remains absent. This work aims to fill this gap by providing a systematic investigation of these aspects, thereby contributing to a more complete understanding of the model.

%%%%%%%%%%%%%%%%%%%%%%%%%%%%%%%%%%%%%%%%%%%%%%%%%%%%%%%%%%%%%%%%%%%%%%%%%%%%%%%%%%%%%%%%%%%%%%%%%%%%%%%%%%%%%%%%%%%%%%%%%%%%%%%%%%%%%%%%%%%%%%%%%%%%%%%%%%%%%%%%%%%%%%%%%%%%%%%%%%%%%%%%%%%%%%%%%%%%%%%%%%%%%%%%%%%%%%%%%%%%%%%%%%%%%%%%%%%%%%%%%%%%%%%%%%%%%%%%%%%%%%%%%%%%%%%%%%%%%%%%%%%%%%%%%%%%%%%%%%%%%%%%%%%%%%%%%%%%%%%%%%%%%%%%%%%%%%%%%%%%%%%%%%%%%%%%%%%%%%

\section{The general setup}
The studies discussed in \cite{zhang2024black} focus on addressing the problem of maintaining covariance within the framework of spherically symmetric vacuum gravity. By preserving the classical theory’s kinematic variables and vector constraints, the authors propose an effective Hamiltonian constraint, $H_{\text{eff}}$, along with a freely defined function that helps build the effective metric. Similar to the classical approach, they assume the existence of a Dirac observable associated with the black hole mass. From this assumption, the authors derive conditions that ensure the preservation of spacetime covariance. These conditions reveal the interdependence of the effective Hamiltonian, the Dirac observable for the black hole mass, and the free function. Solving these conditions produces two distinct families of effective Hamiltonians, each parameterized by a quantum variable. When these quantum parameters approach zero, the classical constraints are recovered. Consequently, two distinct quantum-modified black hole metrics emerge, each corresponding to different spacetime structures.

The quantum--corrected black hole metrics are expressed by the following line element
\begin{equation}\label{metric}
    \mathrm{d}s^2 = -A(r)\mathrm{d}t^2 + \frac{\mathrm{d}r^2}{B(r)} +C(r)(\mathrm{d}\theta^2 + \sin^2\theta \, \mathrm{d}\phi^2).
\end{equation}

Model  \Romannum{1}:
\begin{align}\label{model1}
     A(r) &= f(r)\left(1 + \frac{{{\xi ^2}}}{{{r^2}}}f(r)\right), \hfill \\ \nonumber
  B(r)& = A(r).
\end{align}

Model \Romannum{2}:
\begin{align}\label{model2}
    A(r) &= f(r), \\ \nonumber
  B(r) &= f(r)\left(1 + \frac{{{\xi ^2}}}{{{r^2}}}f(r)\right),
\end{align}
where $f(r)=1-\frac{2M}{r}$, $C(r) = r^2$ and $\xi$ is a quantum parameter, and $M$ represents the ADM mass. The quantum parameter ($ \xi $) in this paper represents a fundamental component in introducing quantum gravitational corrections into the effective Hamiltonian framework. It is proportional to the Planck length ($ \ell_p = \sqrt{\hbar G / c^3} $) and encapsulates the microscopic effects of quantum gravity. Derived through a polymerization procedure, which replaces classical variables with trigonometric functions, the parameter models discrete space-time structures inspired by loop quantum gravity techniques. This approach ensures compatibility with quantum geometry while maintaining general covariance. Mathematically, $ \xi $ governs the oscillatory behavior of connections, such as $ \sin(\xi K) $, effectively capturing curvature fluctuations and non-classical effects in the Hamiltonian formulation.

Physically, the scale of the quantum parameter ($ \xi $) aligns closely with the Planck length, approximately $ 10^{-35} $ meters, indicating that its influence is dominant at microscopic distances near the black hole horizon. The effective metrics derived in the paper demonstrate that modifications proportional to $ \xi^2 / x^2 $ become significant in the vicinity of the horizon but diminish at larger radial distances, preserving classical general relativity at macroscopic scales. Near the event horizon, these corrections prevent singularities by introducing quantum-modified structures resembling double-horizon geometries similar to Reissner-Nordstr\"{o}m black holes. These quantum effects also enables the smooth transitions between black and white hole regions, reinforcing the Planck-scale nature of $ \xi $ \cite{zhang2024black}.

\begin{figure}[ht]
	\centering
	\includegraphics[width=90mm]{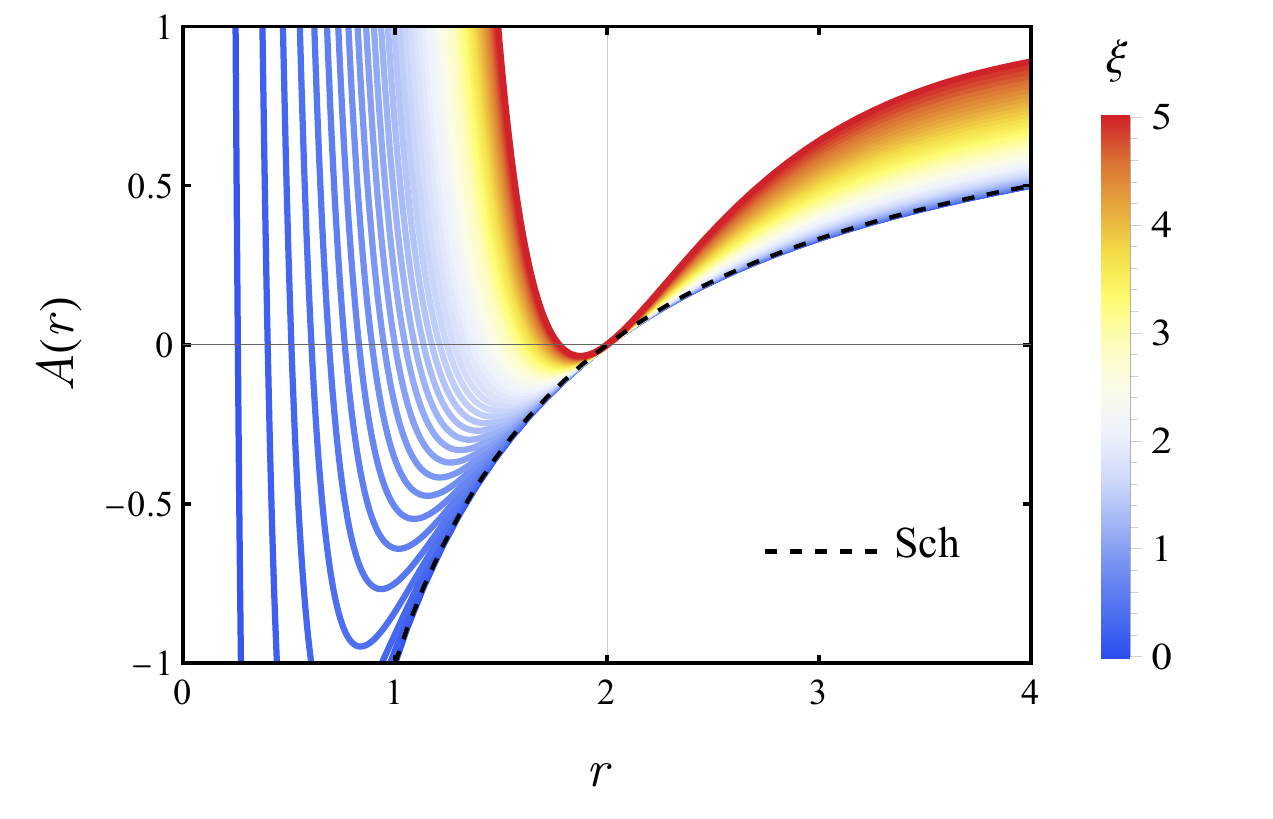}
	\caption{The lapse function of model \Romannum{1}, for various values of $\xi$ is represented for $M = 1$. The dashed line shows the Schwarzschid case}
	\label{fig:horizon}
\end{figure}

In model I, the expression for the event horizon can be found by solving $r$ in $A(r) = 0$. Results give four solutions, where two are imaginary. One of the other two solutions is the horizon for the Schwarzschild case $r_{\rm h} = 2M$, and the other one is given by the exact solution.
\begin{equation} \label{hor}
    r_{\rm h} = \frac{\eta^{1/3}}{3} - \frac{\xi^2}{\eta^{1/3}},
\end{equation}
where we wrote 
\begin{equation}
    \eta = 27 \xi^{2} M +3 \sqrt{81 M^{2} \xi^{4}+3 \xi^{6}},
\end{equation}
for brevity. To study the behavior of the horizon under the influence of the parameter $\xi$, we simplify the expression above using approximations. One of them is when $\xi \rightarrow 0$, Eq. \eqref{hor} becomes
\begin{equation} \label{hor1}
    r_{\rm h} \sim \frac{54^{\frac{1}{3}} M^{\frac{1}{3}} \xi^{\frac{2}{3}}}{3}-\frac{54^{\frac{2}{3}} \xi^{\frac{4}{3}}}{54 M^{\frac{1}{3}}} + \mathcal{O}(\xi^{8/3}),
\end{equation}
which shows that the horizon under the influence of $\xi$ disappears, while the classical horizon still remains. The next approximation to Eq. \eqref{hor} is when $\xi \rightarrow \infty$, which gives
\begin{equation}\label{rhin}
    r_{\rm h} = 2 M -\frac{8 M^{3}}{\xi^{2}}.
\end{equation}
Since the horizon derived in Eq. \eqref{rhin} is smaller than \( 2M \), it lies within the classical horizon, converging to \( 2M \) as \(\xi\) becomes large, therefore Eq. \eqref{rhin} is considered as the inner horizon and thus, we adopt \( r_{\rm h} = 2M \) as the event horizon. In Fig. \ref{fig:horizon}, we present a numerical plot of the horizon for model I, which agrees with our analytical findings.

%%%%%%%%%%%%%%%%%%%%%%%%%%%%%%%%%%%%%%%%%%%%%%%%%%%%%%%%%%%%%%%%%%%%%%%%%%%%%%%%%%%%%%%%%%%%%%%%%%%%%%%%%%%%%%%%%%%%%%%%%%%%%%%%%%%%%%%%%%%%%%%%%%%%%%%%%%%%%%%%%%%%%%%%%%%%%%%%%%%%%%%%%%%%%%%%%%%%%%%%%%%%%%%%%%%%%%%%%%%%%%%%%%%%%%%%%%%%%%%%%%%%%%%%%%%%%%%%%%%%%%%%%%%%%%%%%%%%%%%%%%%%%%%%%%%%%%%%%%%%%%%%%%%%%%%%%%%%%%%%%%%%%%%%%%%%%%%%%%%%%%%%%%%%%%%%%%%%%%%%%%%%%%%%%%%%%%%%%%%%%%%%%%%%%%%%%%%%%%%%%%%%%%%%%%%%%%%%%%%%%%%%%%%%%%%%%%%%%%%%%%%%%%%%%%%

%%%%%%%%%%%%%%%%%%%%%%%%%%%%%%%%%%%%%%%%%%%%%%%%%%%%%%%%%%%%%%%%%%%%%%%%%%%%%%%%%%%%%%%%%%%%%%%%%%%%%%%%%%%%%%%%%%%%%%%%%%%%%%%%%%%%%%%%%%%%%%%%%%%%%%%%%%%%%%%%%%%%%%%%%%%%%%%%%%%%%%%%%%%%%%%%%%%%%%%%%%%%%%%%%%%%%%%%%%%%%%%%%%%%%%%%%%%%%%%%%%%%%%%%%%%%%%%%%%%%%%%%%%%%%%%%%%%%%%%%%%%%%%%%%%%%%%%%%%%%%%%%%%%%%%%%%%%%%%%%%%%%%%%%%%%%%%%%%%%%%%%%%%%%%%%%%%%%%%%%%%%%%%%%%%%%%%%%%%%%%%%%%%%%%%%%%%%%%%%%%%%%%%%%%%%%%%%%%%%%%%%%%%%%%%%%%%%%%%%%%%%%%%%%%%%

%%%%%%%%%%%%%%%%%%%%%%%%%%%%%%%%%%%%%%%%%%%%%%%%%%%%%%%%%%%%%%%%%%%%%%%%%%%%%%%%%%%%%%%%%%%%%%%%%%%%%%%%%%%%%%%%%%%%%%%%%%%%%%%%%%%%%%%%%%%%%%%%%%%%%%%%%%%%%%%%%%%%%%%%%%%%%%%%%%%%%%%%%%%%%%%%%%%%%%%%%%%%%%%%%%
%%%%%%%%%%%%%%%%%%%%%%%%%%%%%%%%%%%%%%%%%%%%%%%%%%%%%%%%%%%%%%%%%%%%%%%%%%%%%%%%%%%%%%%%%%%%%%%%%%%%%%%%%%%%%%%%%%%%%%%%%%%%%%%%%%%%%%%%%%%%%%%%%%%%%%%%%%%%%%%%%%%%%%%%%%%%%%%%%%%%%%%%%%%%%%%%%%%%%%%%%%%%%%%%%%%%%%%%%%%%%%%%%%%%%%%%%%%%%%%%

\section{Partial wave equation}\label{part}

\begin{figure}[ht]
	\centering
	\includegraphics[width=82mm]{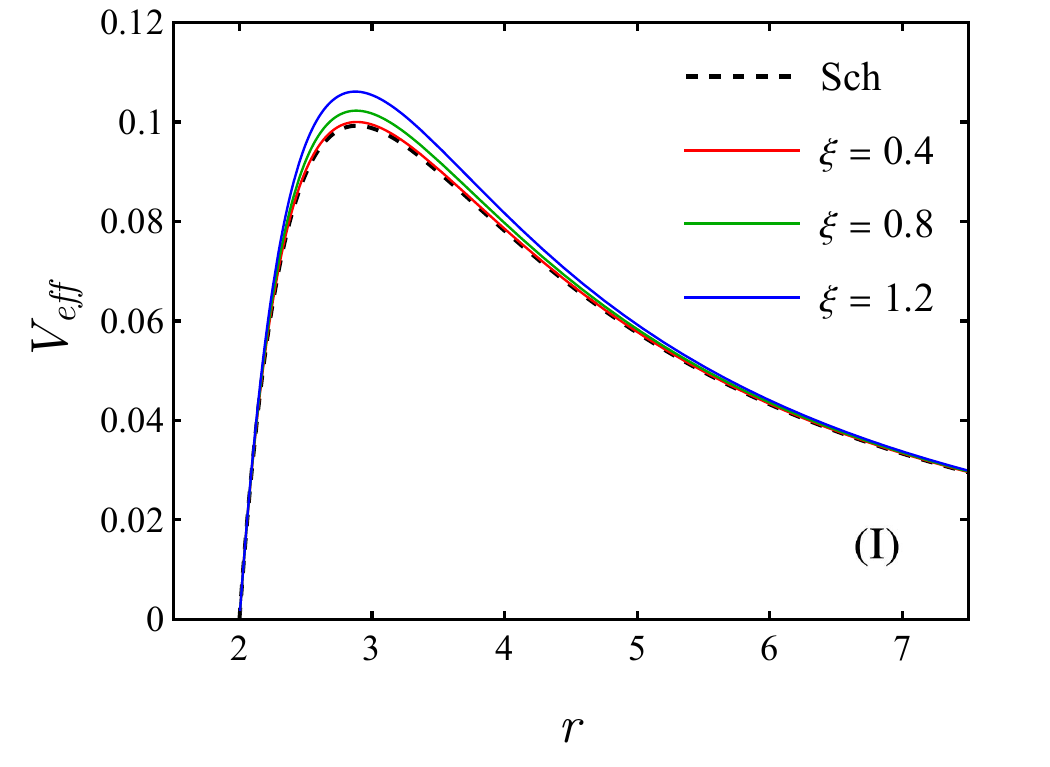}
   \includegraphics[width=80mm]{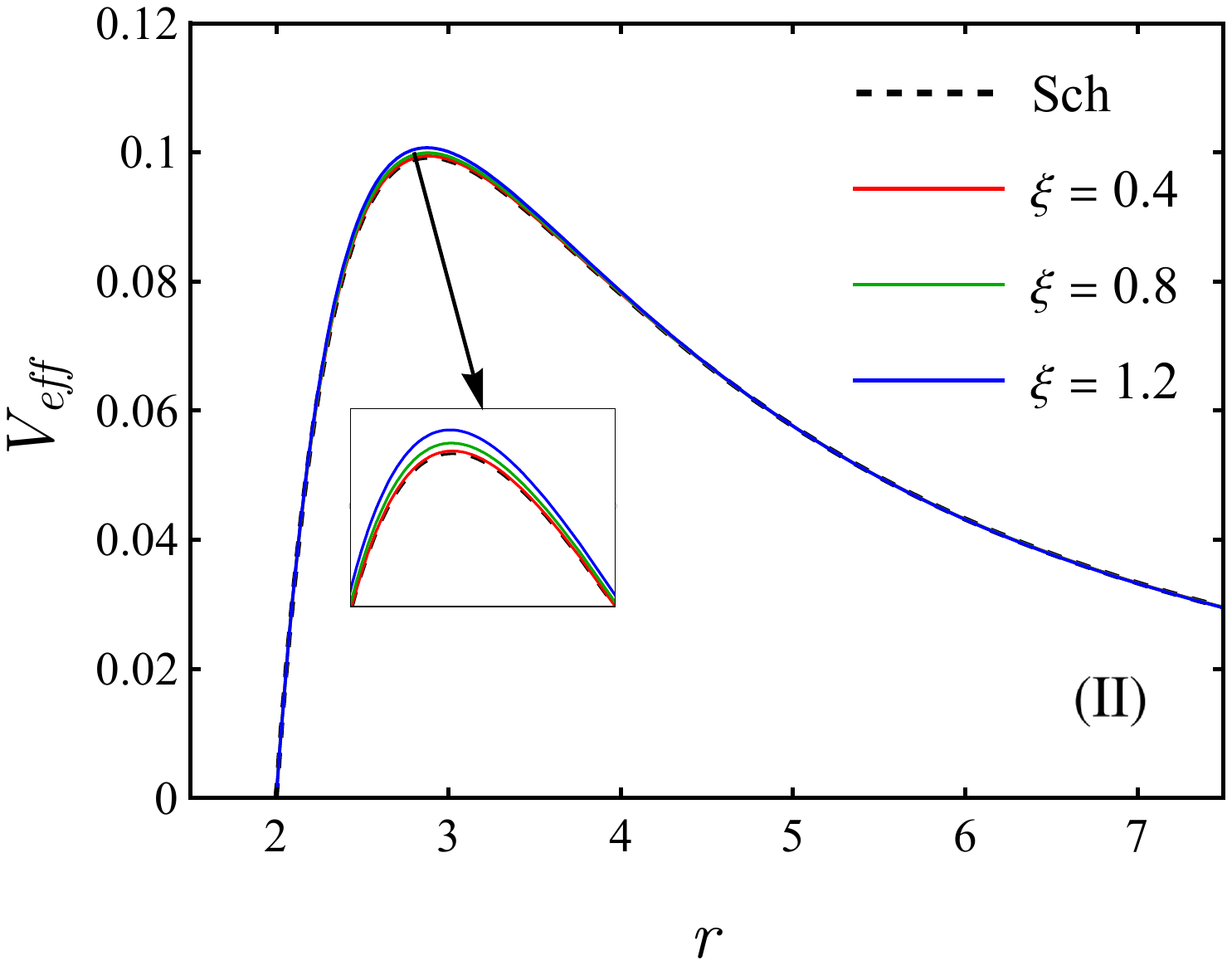}
	\caption{The effective potential is shown for $M = 1$, $l=1$ and different values of $\xi$.}
	\label{fig:Veff}
\end{figure}

This section analyzes the partial wave equation by considering the Klein--Gordon equation in spherically symmetric curved spacetime, previously introduced in Eq. \eqref{model1}-\eqref{model2}

\begin{equation}\label{klein}
	\frac{1}{{\sqrt { - g} }}{\partial _\mu }(\sqrt { - g} {g^{\mu \nu }}{\partial _\nu }\Psi ) = 0.
\end{equation}
By applying the separation of variables $\Psi _{\omega lm}(\mathbf{r},t) = \frac{\psi_{\omega l(r)}}{r}{Y_{lm}}(\theta ,\varphi ){e^{ - i\omega t}}$ \cite{campos2022quasinormal,anacleto2021quasinormal} and also defining the tortoise coordinate ($r^{*}$) through the time and radius metric components as
\begin{equation}\label{rstar}
	{\rm{d}}r^* = \frac{{\rm{d}}r}{\sqrt {A(r)B(r)} },
\end{equation}
the Klein--Gordon equation will be simiplified to a Schr\"{o}dinger--like wave equation
\begin{equation}\label{waves}
	\left[\frac{{{\rm{d}^2}}}{{\rm{d}{{r^*}^2}}} + ({\omega ^2} - {V_{eff}})\right]\psi_{\omega l} (r) = 0.
\end{equation}
Here, \( V_{eff} \) is the effective potential, given by
\begin{equation}
\label{Veff}
V_{eff} = A(r)\left[\frac{{l(l + 1)}}{{{r^2}}} + \frac{1}{{r\sqrt {{A(r)}{B(r)^{ - 1}}} }}\frac{\mathrm{d}}{{\mathrm{d}r}}\sqrt {{A(r)}B(r)}\right].
\end{equation}
To aid understanding, we depict \( V_{eff} \) as a function of the tortoise coordinate considering the metric parameter in Eq. \eqref{model1} and \eqref{model2} for Models \Romannum{1} and \Romannum{2}, respectively. In each plots value of the parameter \( \xi \) varies, including the Schwarzschild case (\( \xi = 0 \)) for both model \Romannum{1} and model \Romannum{2}, in Fig. \ref{fig:Veff}. 

The parameter \( \xi \) affects the effective potential's height. In both models, higher values of \( \xi \) make the height of the \( V_{eff} \) higher. However, in model 2, the effective potential is less sensitive to the variation of \(\xi\).

%%%%%%%%%%%%%%%%%%%%%%%%%%%%%%%%%%%%%%%%%%%%%%%%%%%%%%%%%%%%%%%%%%%%%%%%%%%%%%%%%%%%%%%%%%%%%%%%%%%%%%%%%%%%%%%%%%%%%%%%%%%%%%%%%%%%%%%%%%%%%%%%%%%%%%%%%%%%%%%%%%%%%%%%%%%%%%%%%%%%%%%%%%%%%%%%%%%%%%%%%%%%%%%%%%%%%%%%%%%%%%

\section{Absorption cross--section}\label{abs}

As demonstrated in Fig. \ref{fig:Veff}, the effective potential of the scalar field is concentrated, diminishing to zero at both the horizon and spatial infinity. Given these properties, the asymptotic behavior of the radial wavefunction is expected to be purely incoming at the event horizon, while at spatial infinity it consists of both ingoing and outgoing components. The corresponding boundary conditions are described as follows \cite{macedo2015scattering,macedo2016absorption,anacleto2023absorption}
\begin{equation}\label{bound}
	{\psi _{\omega l}} \approx 
	\begin{cases}
	\mathcal{T}_{\omega l} R_{{1}} , &\text{for} \ {r^*} \to +r_{\rm h} ~ (r \to -\infty )\\
	R_{{2}}+\mathcal{R}_{\omega l} R^*_{{2}},&\text{for} \ {r^*} \to +\infty ~ (r \to \infty ).\\
	\end{cases}
\end{equation}
Here, $|\mathcal{R}_l|^2$ and $|\mathcal{T}_l|^2$ represent the reflection and transmission coefficients, respectively. Conservation of flux ensures that the relationship $|\mathcal{R}_l|^2 + |\mathcal{T}_l|^2 = 1$ holds true. Furthermore, $R_1$ and $R_2$ can be expressed as \cite{macedo2013absorption} 
\begin{align}\label{Rroman1}
	{R_{{1}}}& = {e^{ - i\omega r*}}\sum\limits_{j = 0}^N {{{(r - {r_{\rm h}})}^j}F_{{r_{\rm h}}}^{(j)}}, \\ \label{Rroman2}
	{R_{{2}}}&= {e^{ -i\omega r*}}\sum\limits_{j = 0}^N {\frac{{F_\infty ^{(j)}}}{{{r^j}}}} .
\end{align}
To determine the coefficients $F_\infty^{(j)}$ and $F_{r_{\rm h}}^{(j)}$, it is required that $R_1$ and $R_2$ satisfy the differential equation in Eq. \eqref{waves} in the regions far from the black hole and near the event horizon, respectively. The phase shift, $\delta_{\omega l}$, is defined as \cite{dolan2009scattering}
\begin{equation}\label{phase}
	{e^{2i{\delta _{\omega l}}}} = {( - 1)^{l + 1}}\mathcal{R}_{\omega l}.
\end{equation}	
To calculate the phase shifts for the absorption cross-section, we employed the numerical method discussed in Ref. \cite{crispino2009scattering}, solving the radial equation \eqref{waves}. By using Eqs. \eqref{Rroman1} and \eqref{Rroman2} and evaluating the series up to the 10th order ($N = 10$), we matched the numerical solutions to the boundary conditions in Eq. \eqref{bound}, starting near the horizon at $r/r_{\rm h} - 1 = 10^{-3}$ and extending into the asymptotically flat region at $r \sim 200 r_{\rm h}$. In this manner, the total absorption cross--section is given by $\sigma_{abs} = \sum_{l=0}^\infty \sigma_{abs}^l$, where $\sigma_l$ is defined as \cite{crispino2009scattering}
\begin{equation}\label{partial1}
		\sigma _{abs}^l = \frac{\pi }{{{\omega ^2}}}(2l + 1)(1 - {\left| {{e^{2i{\delta _{\omega l}}}}} \right|^2}).
	\end{equation}
Applying Eq. \eqref{phase} and flux conservation, we get $|e^{2i\delta_{\omega l}}|^2 = 1 - |\mathcal{R}{\omega l}|^2 = |T{\omega l}|^2$ \cite{crispino2013greybody}. Thus, the total absorption cross--section can be expressed as
\begin{equation}\label{partial2}
	{\sigma _{abs}} = \frac{\pi }{{{\omega ^2}}}\sum\limits_{l = 0}^\infty  {(2l + 1)} {\left| {{T_{\omega l}}} \right|^2},
\end{equation}
where $|T_{\omega l}|^2$ is the transmission coefficient, also known as the greybody factor. By inserting the numerical transmission coefficients into Eq. \eqref{partial1}, it can be calculated for various values of $\xi$. In addition, for low frequencies, $\sigma_{abs}$ converges to the area of the black hole's event horizon, as demonstrated in previous works \cite{Das1996we,Higuchi:2001si}.

\begin{figure}[ht!]
	\centering
	\includegraphics[width=80mm]{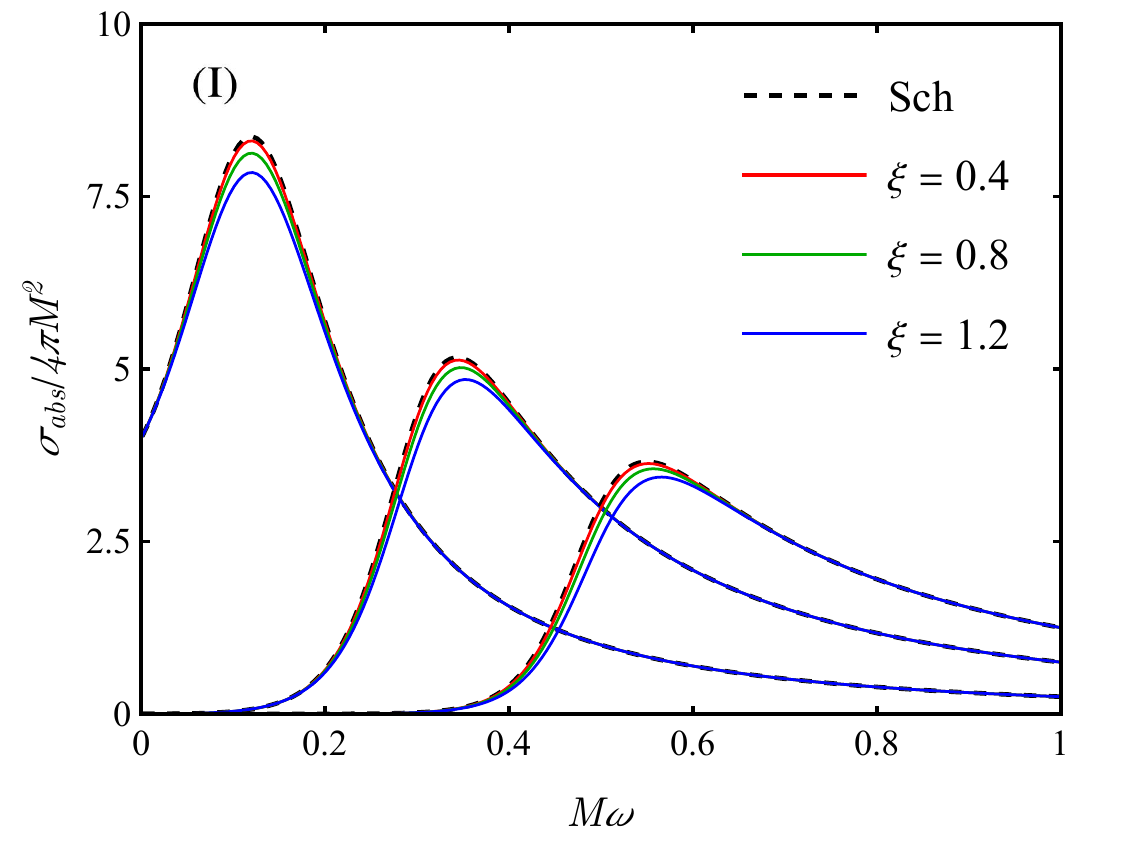}   \includegraphics[width=80mm]{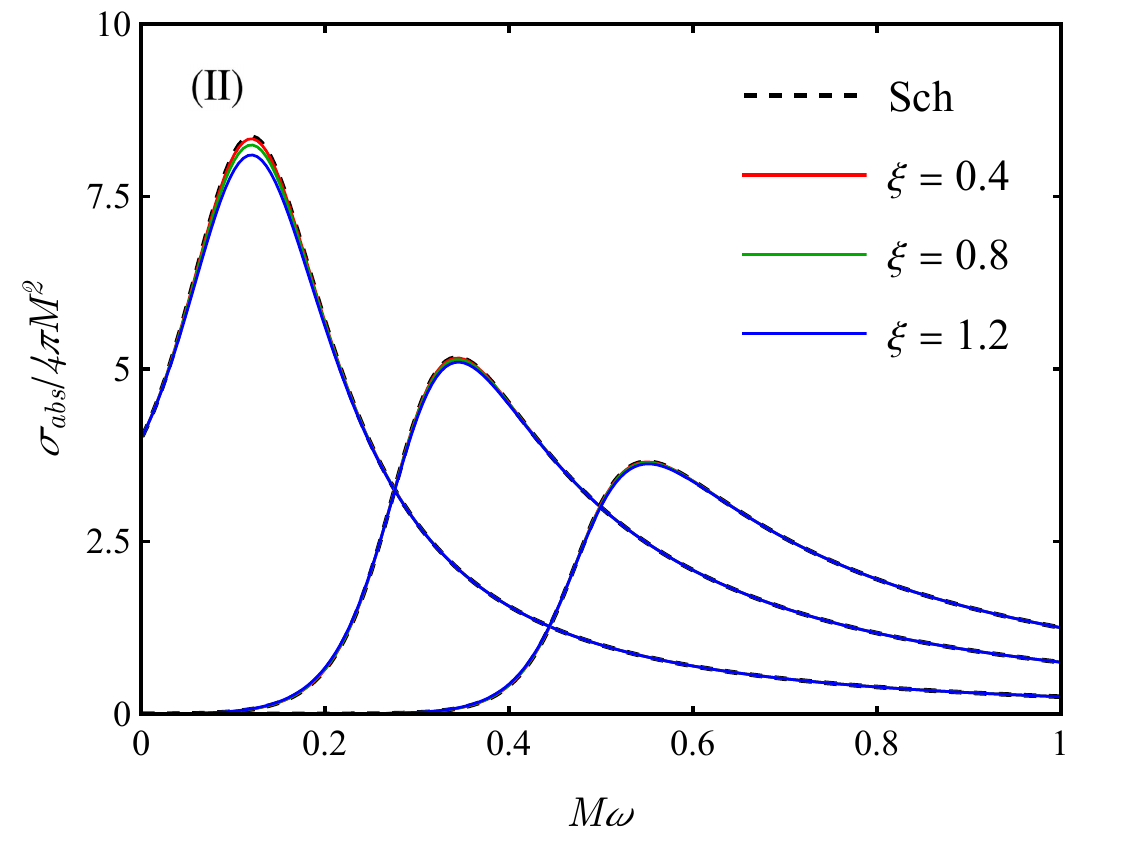} 
	\caption{Partial absorption cross--sections for $l=0,1,2$ and different choices of parameter $\xi$.}
	\label{fig:Psigma}
\end{figure}

%%%%%%%%%%%%%%%%%%%%%%%%%%%%%%%%%%%%%%%%%%%%%%%%%%%%%%%%%%%%%%%%%%%%%%%%%%%%%%%%%%%%%%%%%%%%%%%%%%%%%%%%%%%%%%%%%%%%%%%%%%%%%%%%%%%%%%%%%%%%%%%%%%%%%%%%%%%%%%%%%%%%%%%%%%%%%%%%%%%%%%%%%%%%%%%%%%%%%%%%%%%%%%%%%%%%%%%%%%%%%%%%%%%%%%%%%%%%%%%%%%%%%%%%%%%%%%%%%%%%%%%%%%%%%%%%%%%%%%%%%%%%%%%%%%%%%%%%%%%%%%%%%%%%%%%%%%%%%%%%%%%%%%%%%%%%%%%%%%%%%%%%%%%%%%%%%%%%%%%%%%%%%%%%%%%%%%%%%%%%%%%%%%%%%%%%%%%%%%%%%%%%%%%%%%%%%%%%%%

\subsection{Low--Frequency Regime}

At low frequencies, the absorption cross--section for a black hole typically approaches the area of the event horizon. This behavior is well--documented across various studies and is a fundamental feature of black hole physics\cite{Das1996we, Higuchi:2001si}. As the incident wave frequency decreases, the interaction between the wave and the black hole becomes dominated by long--wavelength effects, causing the absorption cross--section to settle toward the value given by the black hole's surface area, particularly at the event horizon.

This phenomenon can be attributed to the fact that, at low frequencies, the wave's wavelength becomes comparable to or larger than the size of the black hole, leading to a simplified interaction where the cross--section reflects the fundamental size of the black hole itself. Specifically, in the case of scalar field perturbations, the low--frequency absorption cross--section for a Schwarzschild black hole is shown to approach the geometrical area of the horizon asymptotically. This limit holds across different scenarios, including modifications in the black hole's parameters, rotating black holes, or surrounding spacetime \cite{leite2017scalar,macedo2013absorption}, as long as the frequency remains sufficiently low.

In the case of effective quantum gravity, the event horizon in both cases of Models \Romannum{1} and \Romannum{2}, is $r_{\rm h}=2M$, therefore, we expect the absorption cross--section at low--frequency regime to be $\sigma\approx 4\pi r_{\rm h}^2=16\pi M^2$. We can check this parameter in Fig. \ref{fig:Psigma}. As expected, when frequency tends to zero, $\sigma/4\pi M^2\approx 4$, which is equal to $\sigma_{abs}$ at the event horizon.

%%%%%%%%%%%%%%%%%%%%%%%%%%%%%%%%%%%%%%%%%%%%%%%%%%%%%%%%%%%%%%%%%%%%%%%%%%%%%%%%%%%%%%%%%%%%%%%%%%%%%%%%%%%%%%%%%%%%%%%%%%%%%%%%%%%%%%%%%%%%%%%%%%%%%%%%%%%%%%%%%%%%%%%%%%%%%%%%%%%%%%%%%%%%%%%%%%%%%%%%%%%%%%%%%%%%%%%%%%%%%%%%%%%%%%%%%%%%%%%%%%%%%%%%%%%%%%%%%%%%%%%%%%%%%%%%%%%%%%%%%%%%%%%%%%%%%%%%%%%%%%%%%%%%%%%%%%%%%%%%%%%%%%%%%%%%%%%%%%%%%%%%%%%%%%%%%%%%%%%%%%%%%%%%%%%%%%%%%%%%%%%%%%%%%%%%%%%%%%%%%%%%%%%%%%%%%%%%%%

\subsection{High--Frequency Regime}

In this section, we analyze the high--frequency limit of the absorption cross--section, also known as the geometric cross--section \(\sigma_{\text{geo}}\). In this regard, it can be expressed as 
\begin{equation}\label{eq:geocross}
    \sigma_{\text{geo}} = \pi b_{\rm{c}}^2,
\end{equation}
where \(b_c\) represents the critical impact parameter for a light ray approaching the black hole. 
Stated differently, $\sigma_{abs}$ of null geodesics tends to the captured cross--section in the high--frequency limit.
We apply the critical impact parameter $b_\rm{c}=3\sqrt{3}M$ for the second model and for the first model according to Eq. \eqref{bcrit} and arrive at two geometric cross--section for model \Romannum{1} and \Romannum{2}.
\begin{align}\label{geo1}
 \sigma_{\text{geo}}^{\Romannum{1}}&=\frac{729\pi M^4}{\left(27M^2+\xi^2\right)},\\  \label{geo2} 
 \sigma_{\text{geo}}^{\Romannum{2}}&=27\pi M^2.
\end{align}

\begin{figure}[ht!]
    \centering
    \includegraphics[width=80mm]{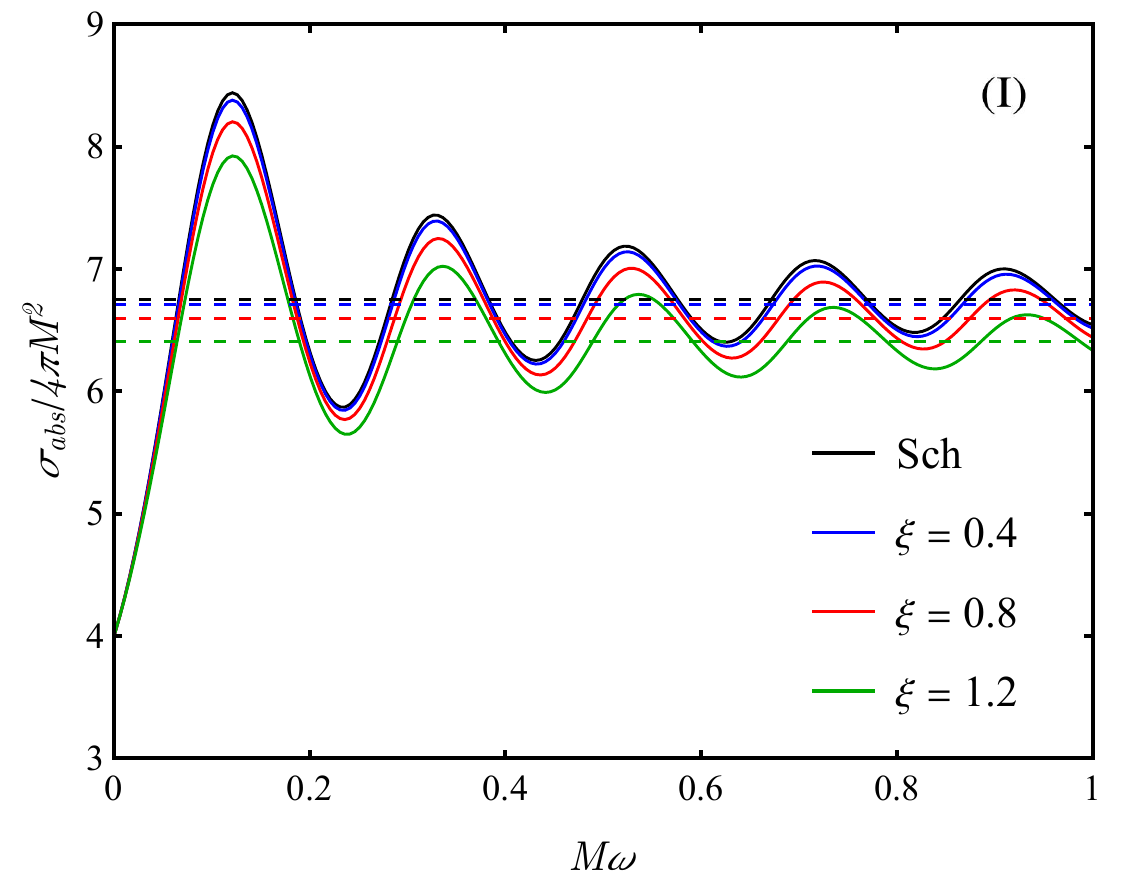}
    \includegraphics[width=80mm]{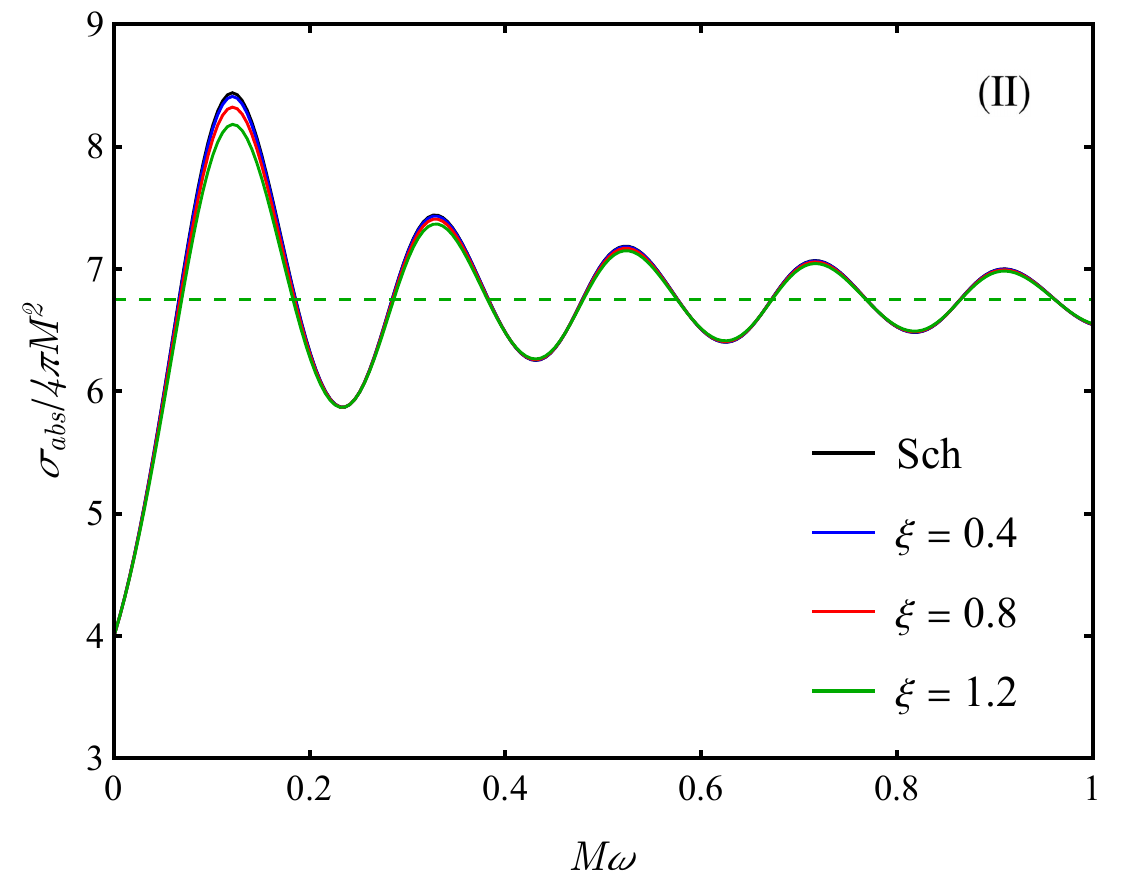}
    \caption{Total absorption cross--sections for \(l=0\) to \(l=6\) and different values of \(\xi\). The dashed lines represent the geometrical cross--sections.}
    \label{fig:Tsigma}
\end{figure}
Fig. \ref{fig:Tsigma} illustrates the total absorption cross--section for \(l = 0\) to \(l = 6\) and varying \(\xi\). 
In these figures, the geometric cross--section obtained via Eq. \eqref{eq:geocross}, is represented by the dashed horizontal lines. It is noticeable that in such an approximation, it converges to a constant value that is consistent with \(\sigma_{\text{geo}}\) derived in Eq. \eqref{geo1} and \eqref{geo2}.
In addition, in model II, the geometric cross--section does not depend on the $\xi$.

%%%%%%%%%%%%%%%%%%%%%%%%%%%%%%%%%%%%%%%%%%%%%%%%%%%%%%%%%%%%%%%%%%%%%%%%%%%%%%%%%%%%%%%%%%%%%%%%%%%%%%%%%%%%%%%%%%%%%%%%%%%%%%%%%%%%%%%%%%%%%%%%%%%%%%%%%%%%%%%%%%%%%%%%%%%%%%%%%%%%%%%%%%%%%%%%%%%%%%%%%%%%%%%%%%%%%%%%%%%%%%%%%%%%%%%%%%%%%%%%%%%%%%%%%%%%%%%%%%%%%%%%%%%%%%%%%%%%%%%%%%%%%%%%%%%%%%%%%%%%%%%%%%%%%%%%%%%%%%%%%%%%%%%%%%%%%%%%%%%%%%%%%%%%%%%%%%%%%%%%%%%%%%%%%%%%%%%%%%%%%%%%%%%%%%%%%%%%%%%%%%%%%%%%%%%%%%%%%%%%%%%%%%%%%

\section{Greybody factors}\label{Tbound}

Black holes emit radiation, known as \textit{Hawking} radiation, due to quantum effects near the event horizon. As this radiation moves away from the event horizon, it interacts with the curved spacetime around the black hole, resulting in changes to both the spectrum and intensity of the \textit{Hawking} radiation reaching infinity. Thus, an observer at an infinite distance will detect it in a modified manner. The degree to which this modified spectrum deviates from a perfect black body spectrum is measured by a parameter called the greybody factor. In this section, we examine it for a massless spin--0 and spin--1/2 particles emitted from quantum--modified black holes by effective quantum gravity, using general semi--analytic estimates \cite{sakalli2022topical,boonserm2019greybody,ovgun2024shadow,al2024fermionic,boonserm2008transmission}. The precise bound for the greybody factor, denoted as $T_b$, is given by
\begin{equation}\label{Tb0}
{T_b} \ge {\mathop{\rm sech}\nolimits} ^2 \left(\int_\infty^ {+\infty} {\mathcal{G} \rm{d}}r^{*}\right),
\end{equation}
where 
\begin{equation}\label{Tb00}
\mathcal{G} = \frac{{\sqrt {{{(h')}^2} + {{({\omega ^2} - {V_{eff}} - {h^2})}^2}} }}{{2h}}.
\end{equation}
Here, $h$ is a positive function that satisfies the conditions $h(+\infty) = h(-\infty) = \omega$ . By setting $h$ equal to $\omega$, Equation \eqref{Tb1} simplifies to
\begin{equation}\label{Tb000}
{T_b} \ge {\mathop{\rm sech}\nolimits} ^2 \left(\int_{-\infty}^ {+\infty} \frac{V_{eff}} {2\omega}\rm{d}r^{*}\right)={\mathop{\rm sech}\nolimits} ^2 \left(\int_{r_{\rm h}}^ {+\infty} \frac{V_{eff}} {2\omega\sqrt{A(r)B(r)}}\rm{d}r\right).
\end{equation}

%%%%%%%%%%%%%%%%%%%%%%%%%%%%%%%%%%%%%%%%%%%%%%%%%%%%%%%%%%%%%%%%%%%%%%%%%%%%%%%%%%%%%%%%%%%%%%%%%%%%%%%%%%%%%%%%%%%%%%%%%%%%%%%%%%%%%%%%%%%%%%%%%%%%%%%%%%%%%%%%%%%%%%%%%%%%%%%%%%%%%%%%%%%%%%%%%%%%%%%%%%%%%%%%%%%%%%%%%%%%%%%%%%%%%%%%%%%%%%%%%%%%%%%%%%%%%%%%%%%%%%%%%%%%%%%%%%%%%%%%%%%%%%%%%%%%%%%%%%%%%%%%
\subsection{Greybody factor for bosons}

In this section, we will obtain the greybody factor of the massless scalar particle with spin $0$, by applying the effective potential calculated in Eq.\eqref{Veff} for both models. After substituting the effective potentials  arrive at the following expressions for models \Romannum{1} and \Romannum{2}, respectively
\begin{equation}\label{Tb1}
	T^{\Romannum{1}}_b \ge {\mathop{\rm sech}\nolimits}^2 {\left(\int_{2M}^ {+\infty} \frac{r^3 (2 M+r l  (l +1))-2 \xi ^2 (r-4 M) (r-2 M)}{r^6} \rm{d}r\right) },
\end{equation}
and 
\begin{align}\label{Tb2}
   T^{\Romannum{2}}_b \ge {\mathop{\rm sech}\nolimits}^2 &\Biggl( \int_{2M}^ {+\infty}\frac{\sqrt{(r-2 M) \left(\xi ^2 (r-2 M)+r^3\right)} }{r^8} \\ \nonumber
    & \left(r^3 (2 M+r \ell  (\ell +1))-\xi ^2 (r-5 M) (r-2 M)\right) \rm{d}r \Biggl).
\end{align}

\begin{figure}[ht!]
    \centering
    \includegraphics[width=80mm]{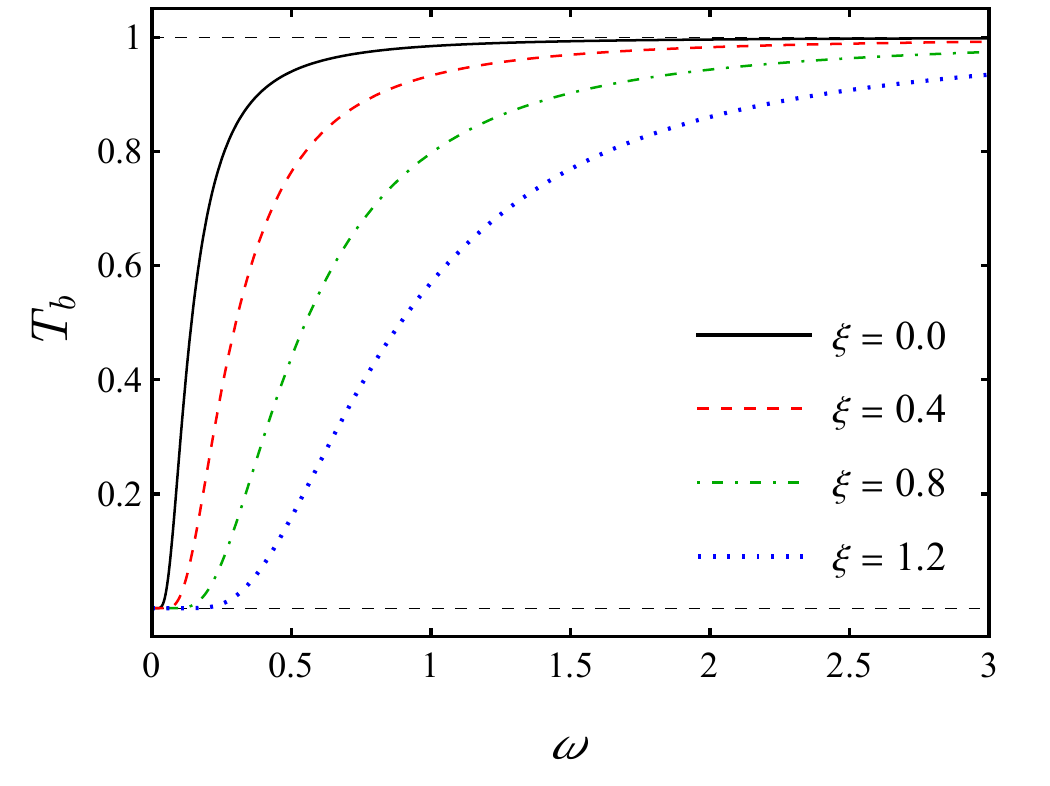}
    \includegraphics[width=80mm]{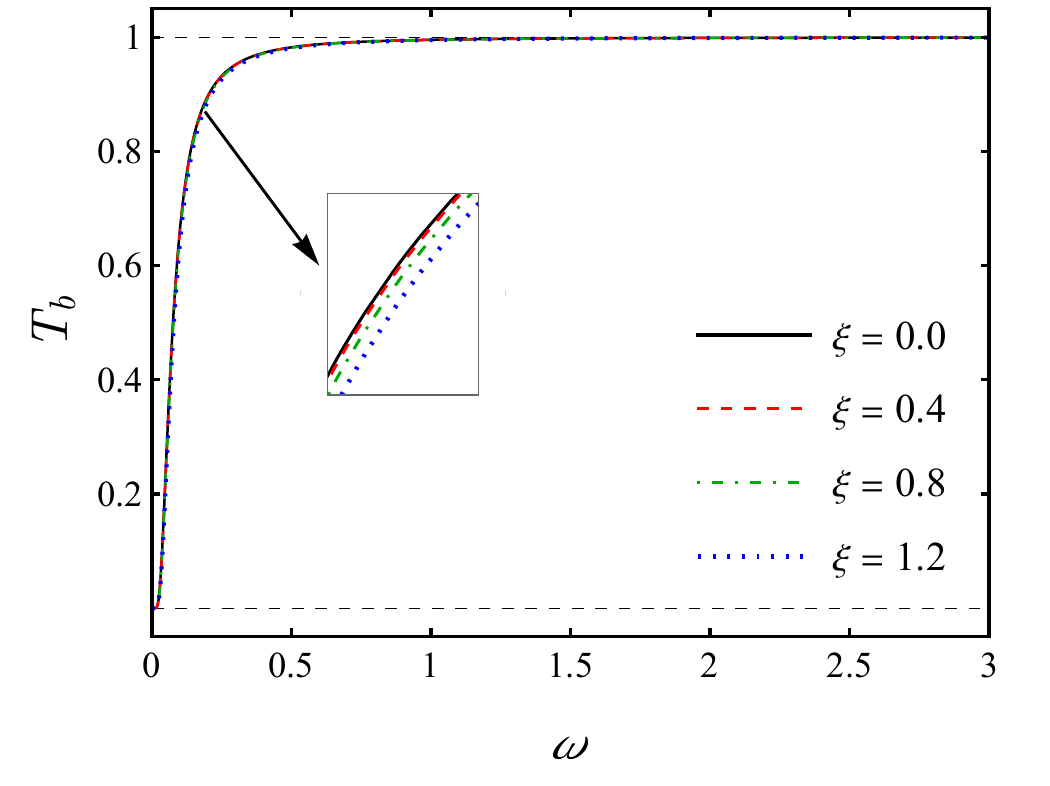}
    \caption{The analysis of the greybody factor associated with scalar fields is presented for \(M = 1\), $l = 1$ alongside a range of \(\xi\) values. The left panel and right panel are allocated to model \Romannum{1} and model \Romannum{2}, respectively.}
    \label{fig:GBF1}
\end{figure}

According to Eq. \eqref{Tb1} and \eqref{Tb2}, $\xi$ plays a significant role in the behavior of the greybody factor. In Fig.  \eqref{fig:GBF1}, we can see it is zero at very low frequencies but rises to $1$ as the frequency gets higher. At lower frequencies, the wave is completely reflected by the potential barrier. However, as the frequency increases, the wave partially passes through, due to a quantum phenomenon called tunneling. When the frequency reaches a certain critical point, the reflection stops entirely. The Fig. \eqref{fig:GBF1} also demonstrates that the greybody factors tend to decrease with higher values of$\xi$ which means that as $\xi$ increases, more of the incoming wave gets scattered which is in consistency with the behavior of $V_{eff}$ in Fig. \ref{fig:Veff}. When $\xi$ increases, the potential barrier also increases, reducing the chance of the wave being transmitted.

Although its behavior in response to an increase in the $\xi$ values is the same in Models I and II, it is noticeable from Fig. \ref{fig:GBF1} that model II is less sensitive to variations in the effective quantum gravity parameter and shows fewer changes from the usual Schwarzschild case. It is worth mentioning that a similar analysis has been accomplished in the literature in the context of other modified theories of gravity \cite{araujo2024exploring,heidari2024impact,heidari2024exploring,heidari2024scattering,hosseinifar2024shadows,chen2024thermal}.

\begin{figure}[ht!]
    \centering
    \includegraphics[width=80mm]{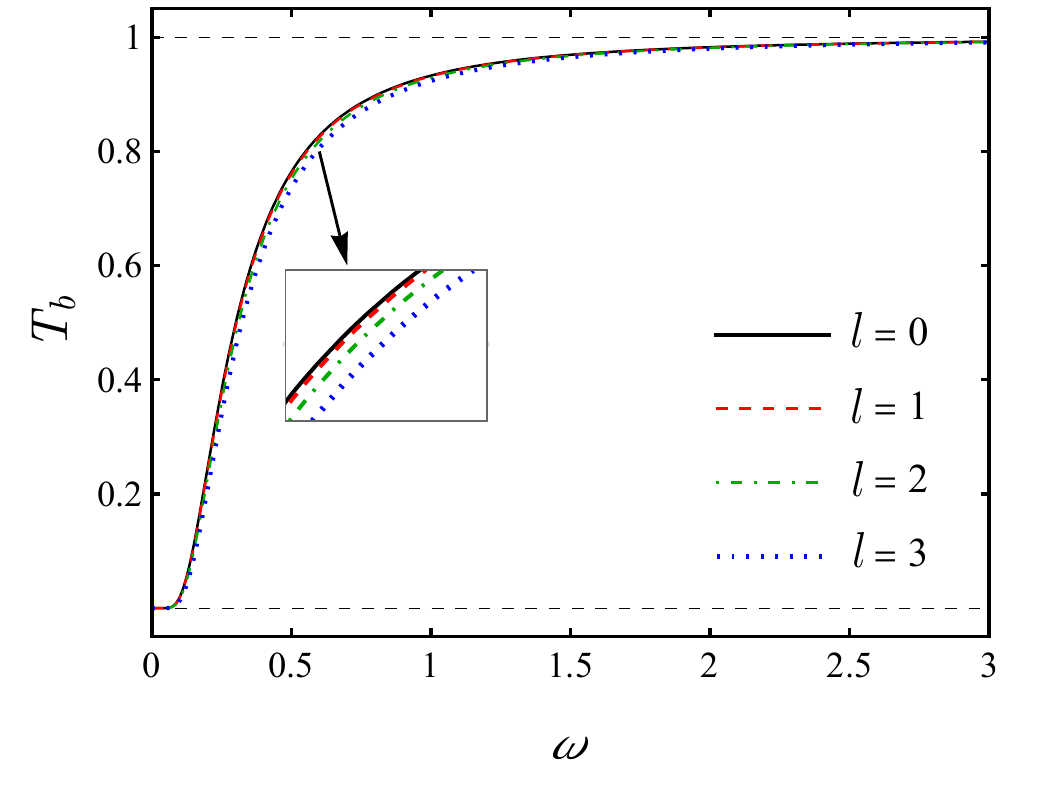}
    \includegraphics[width=80mm]{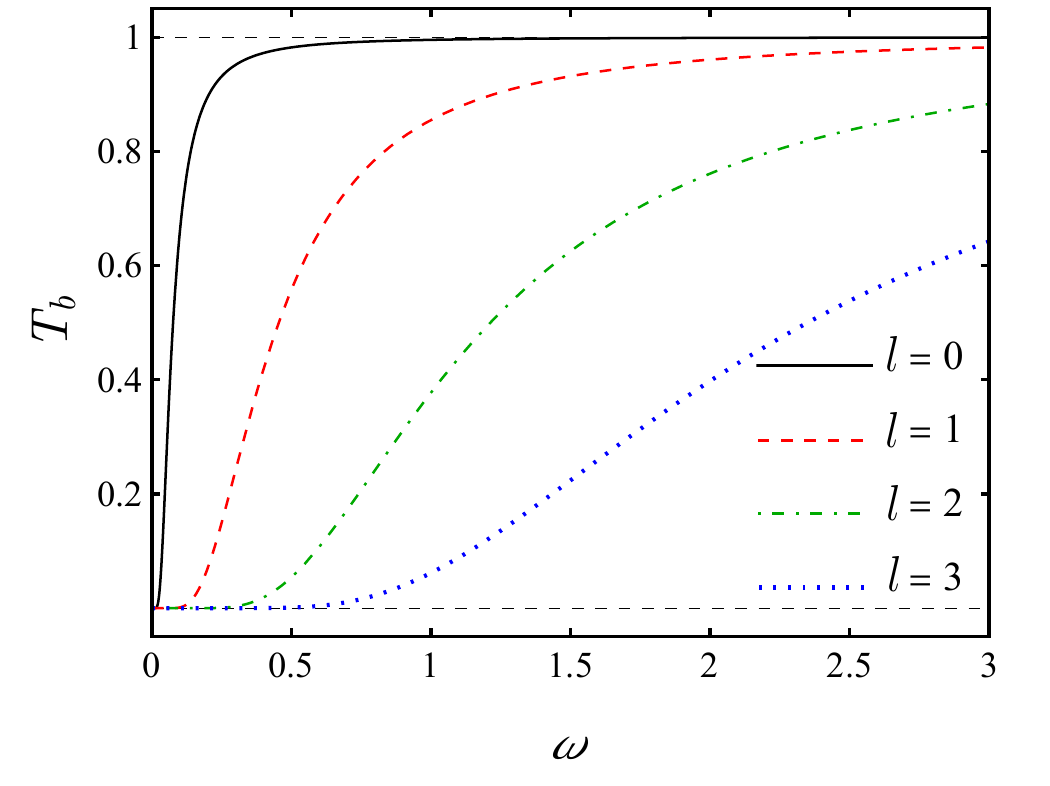}
    \caption{The comparison between greybody factor of scalar field are shown for $M = 1$ and various values of \(l\). The left panel and right panel are devoted to model \Romannum{1} and \Romannum{2}, respectively.}
    \label{fig:GBF2}
\end{figure}

On the other hand, Fig. \ref{fig:GBF2} represents it concerning frequency for a fixed value of $\xi$ and different \(l\). The left and right panels are devoted to model \Romannum{1} and \Romannum{2}, respectively. It shows that in both models the wave would be scattered less from the potential barrier when the $l$ increases. However, the transmission is less sensitive to variation $l$ in model \Romannum{1}.   

%%%%%%%%%%%%%%%%%%%%%%%%%%%%%%%%%%%%%%%%%%%%%%%%%%%%%%%%%%%%%%%%%%%%%%%%%%%%%%%%%%%%%%%%%%%%%%%%%%%%%%%%%%%%%%%%%%%%%%%%%%%%%%%%%%%%%%%%%%%%%%%%%%%%%%%%%%%%%%%%%%%%%%%%%%%%%%%%%%%%%%%%%%%%%%%%%%%%%%%%%%%%%%%%%%%%%%%%%%%%%%%%%%%%%%%%%%%%%%%%%%%%%%%%%%%%%%%%%%%%%%%%%%%%%%%%%%%%%%%%%%%%%%%%%%%%%%%%%%%%%%%%

\subsection{Greybody factor for fermions}

In this section, we examine the massless Dirac perturbation within the context of a static, spherically symmetric black hole in effective quantum gravity spacetime. We utilize the Newman--Penrose formalism to analyze the dynamics of the massless spin--1/2 field. The Dirac equations can be expressed as follows \cite{newman1962approach,chandrasekhar1984mathematical}
\begin{align}
(D + \epsilon - \rho) F_1 +( \bar{\delta} + \pi - \alpha) F_2 &= 0, \\
(\Delta + \mu - \gamma) F_2 + (\delta + \beta - \tau) F_1 &= 0,
\end{align}

where $F_1$ and $F_2$ denote the Dirac spinors, with $D = l^\mu \partial_\mu$, $\Delta = n^\mu \partial_\mu$, $\delta = m^\mu \partial_\mu$, and $\bar{\delta} = \bar{m}^\mu \partial_\mu$ representing the directional derivatives associated with the chosen null tetrad.

In this study, the appropriate null tetrad basis vectors, represented in terms of the metric elements (for simplicity $A(r)$ and $B(r)$ are written as $A$ and $B$ in the following calculations), are given by
\begin{align}
l^\mu &= \left(\frac{1}{A}, \sqrt{\frac{B}{A}}, 0, 0\right), \quad \quad
n^\mu = \frac{1}{2} \left(1, -\sqrt{A B}, 0, 0\right), \\
m^\mu &= \frac{1}{\sqrt{2} r} \left(0, 0, 1, \frac{i}{\sin \theta}\right), \quad \quad
\bar{m}^\mu = \frac{1}{\sqrt{2} r} (0, 0, 1, \frac{-i}{\sin \theta}).
\end{align}

From these definitions, we find that the non--zero components of the spin coefficients are:
\begin{equation}
 \rho = -\frac{1}{r} \frac{B}{A}, \quad \quad
\mu = -\frac{\sqrt{A B}}{2r},  \quad \quad
\gamma = \frac{A'}{4}\sqrt{\frac{B}{A}}, \quad \quad
\beta = -\alpha = \frac{\cot{\theta}}{2\sqrt{2}r} .
\end{equation}

A single equation of motion for $F_1$ is derived by decoupling the equations, that describe the behavior of a massless Dirac field
\begin{align}
\left[(D - 2\rho)(\Delta + \mu - \gamma) - (\delta + \beta) (\bar{\delta}+\beta)\right] F_1 = 0.
\end{align}

Substituting the expressions for the directional derivatives and spin coefficients allows us to rewrite this equation explicitly as follows
\begin{align}
&\left[ \frac{1}{2A} \partial_t^2 - \left( \frac{\sqrt{AB}}{2r} +\frac{A'}{4}\sqrt{\frac{B}{A}}\right)\frac{1}{A}\partial_t- \frac{\sqrt{AB}}{2} \sqrt{\frac{B}{A}}\partial_r^2 -\sqrt{\frac{B}{A}} \partial_r \left( \frac{\sqrt{AB}}{2} + \frac{A'}{4}{\sqrt{\frac{B}{A}}} \right) \right] F_1 +\\ \nonumber
&\left[ \frac{1}{\sin^2\theta} \partial_\phi^2 + i \frac{\cot \theta}{\sin \theta}\partial_\phi + \frac{1}{\sin \theta}\partial_\theta \left( \sin \theta \partial_\theta \right) - \frac{1}{4} \cot^2 \theta + \frac{1}{2} \right] F_1 = 0.
\end{align}

\begin{figure}[ht!]
    \centering
    \includegraphics[width=80mm]{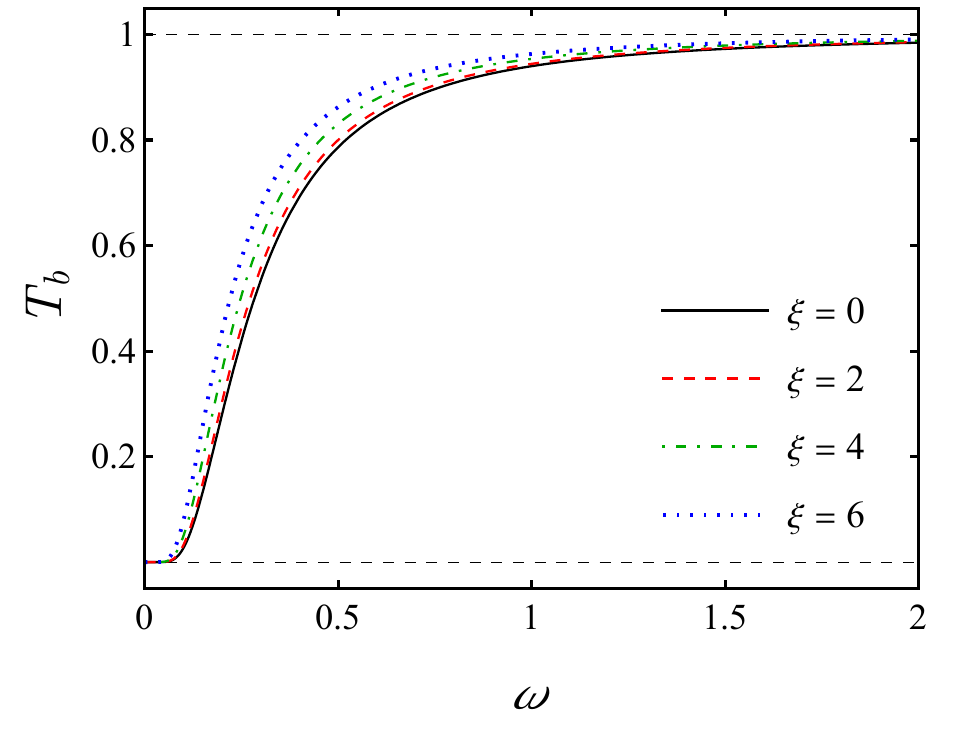}
    \includegraphics[width=80mm]{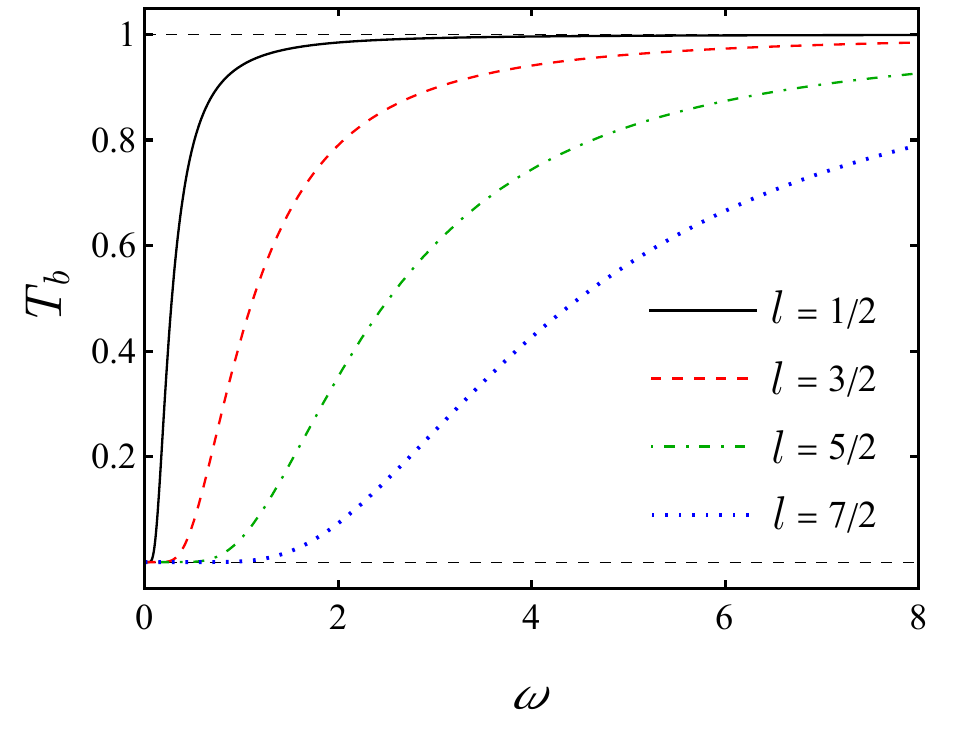}
    \caption{The greybody factor bounding of the massless Dirac field for model \Romannum{2}: In the left panel the greybody factor is plotted for $M = 1$, $l = 1$, and various values of $\xi$ and in the right panel the greybody bounding is shown for $M = 1$, $\xi = 1$ and $l = 1/2$ to $7/2$.}
    \label{fig:GBF3}
\end{figure}

To separate the equations into radial and angular components, we assume the following form for the wave function:
\begin{align}
F_1 = R(r) S_{l,m}(\theta, \phi) e^{-i \omega t},
\end{align}
where the radial component will be formulated as follows
\begin{align}
\left[\frac{-\omega^2}{2A} - \left(\frac{\sqrt{AB}}{2r}+\frac{A'}{4} + \sqrt{\frac{B}{A}}\right)\frac{- i\omega}{A} - \frac{\sqrt{AB}}{2} \sqrt{\frac{B}{A}}\partial_r^2 - \sqrt{\frac{B}{A}}\partial_r \left(\frac{\sqrt{AB}}{2r} + \frac{A'}{4}\sqrt{\frac{B}{A}}\right)-\lambda_l\right] R(r) = 0.
\end{align}
In this case, $\lambda_l$ serves as the separation constant.
Applying the generalized tortoise coordinate $r^*$ is defined in Eq. \eqref{rstar}, the radial wave equation can be transformed into a Schr{\"o}dinger--like wave equation represented as
\begin{align}
\left[\frac{\mathrm{d}^2 }{\mathrm{d}r_*^2} +( \omega^2 - V_{eff}^{\pm}) \right]U_{\pm} = 0,
\end{align}
and the potentials $V_{eff}^{\pm}$ for the massless spin-1/2 field are given by \cite{albuquerque2023massless, al2024massless,arbey2021hawking}

%\begin{align}
%V_{eff}^{\pm} = \frac{(l + \frac{1}{2})^2}{r^2} A \pm \frac{(l + \frac{1}{2})}{r} \sqrt{A B} \left(\frac{A'}{2\sqrt{A}} -\frac{\sqrt{A}}{r}\right).
%\end{align}

\begin{align}\label{Vpm}
V_{eff}^{\pm} = \frac{(l + \frac{1}{2})^2}{r^2} A(r) \pm (l + \frac{1}{2}) \sqrt{A(r) B(r)} \partial_r \left(\frac{\sqrt{A(r)}}{r}\right).
\end{align}
We shall select this potential as $V^+$ without any loss of generality. A similar analysis can be conducted for $V^-$; since the behavior of $V^-$ is qualitatively comparable to that of $V^+$  \cite{albuquerque2023massless,devi2020quasinormal}, we will focus on $V^+$ moving forward.
Now, we substitute the Dirac effective potential from Eq. \eqref{Vpm} in Eq. \eqref{Tb000}, and the greybody factor bounds for both Models are derived. In model \Romannum{1}, the greybody factor can be simplified to 

\begin{align}\label{Tb3}
	T^{\Romannum{1}}_b \ge &{\mathop{\rm  sech}\nolimits}^2 {\left(\frac{1}{2\omega}\int_{2M}^ {+\infty} \frac{V_{eff}^{+}} {\sqrt{A(r)B(r)}} \rm{d}r\right) }\\
 &={\mathop{\rm  sech}\nolimits}^2 {\left(\frac{1}{2\omega}\int_{2M}^ {+\infty}
 \left[\frac{\left(l+\frac{1}{2}\right)^2}{r^2}+\left(l+\frac{1}{2}\right) \partial_r\frac{\sqrt{A(r)}}{r}\right] \rm{d}r\right) } ={\mathop{\rm  sech}\nolimits}^2\left(\frac{(l+\frac{1}{2})^2}{4M\omega}\right).
\end{align}
From Eq. \eqref{Tb3}, it is notable that the result of the greybody factor does not depend on the $\xi$ values. Therefore the influence of effective quantum gravity in Dirac perturbation of model \Romannum{1} is not distinguishable.
On the other hand, the greybody factor for model \Romannum{2} results in 

\begin{align}\label{Tb4}
	T^{\Romannum{2}}_b \ge &{\mathop{\rm  sech}\nolimits}^2 {\left(\frac{1}{2\omega}\int_{2M}^ {+\infty} \frac{V_{eff}^{+}} {\sqrt{A(r)B(r)}} \rm{d}r\right) }\\
 &={\mathop{\rm  sech}\nolimits}^2 {\left(\frac{1}{2\omega}\int_{2M}^ {+\infty} 
 \left[\frac{\left(l+\frac{1}{2}\right)^2}{r^2}\sqrt{\frac{A(r)}{B(r)}}+\left(l+\frac{1}{2}\right) \partial_r\frac{\sqrt{A(r)}}{r}\right] \rm{d}r\right) }\\
 &={\mathop{\rm  sech}\nolimits}^2 {\left(\frac{1}{2\omega}\int_{2M}^ {+\infty} 
 \frac{\left(l+\frac{1}{2}\right)^2}{\sqrt{r^4+\xi ^2 r (r-2 M)}} \rm{d}r\right) }.
\end{align}
Fig. \ref{fig:GBF3}, the variation of the greybody factor concerning frequency is depicted. In the left panel, the behavior of greybody factor is shown for various values of $\xi$ with fixed values of $M = 1$ and $l = 1$. The plots represent that when the effective quantum gravity parameter goes higher, it increases as well, which means that more transmission occurs. In the right panel, it is shown for different angular modes $l \ge s$. It is noticeable that for higher $l$, the wavefunction is less probable to transmit.

%%%%%%%%%%%%%%%%%%%%%%%%%%%%%%%%%%%%%%%%%%%%%%%%%%%%%%%%%%%%%%%%%%%%%%%%%%%%%%

%%%%%%%%%%%%%%%%%%%%%%%%%%%%%%%%%%%%%%%%%%%%%%%%%%%%%%%%%%%%%%%%%%%%%%%%%%%%%%%%%%%%%%%%%%%%%%%%%%%%%%%%%%%%%%%%%%%%%%%%%%%%%%%%%%%%
\section{Geodesics}

%\subsection{Model I}
Geodesics play a crucial role in understanding fundamental physics, as they provide insights into spacetime curvature and describe the motion of particles in the presence of gravitational fields. Studying geodesics within the framework of effective quantum gravity has become a rapidly growing research area, focusing on how quantum corrections influence spacetime properties. By analyzing geodesics in these settings, one can better comprehend the dynamics of particles and fields at the microscopic scale, where quantum phenomena are significant. In this section, we aim to carry out a detailed exploration of geodesic behavior under such conditions. The geodesic equation is given by
\ie
\frac{\mathrm{d}^{2}x^{\mu}}{\mathrm{d}s^{2}} + \Gamma^\mu_{\alpha \beta} \frac{\mathrm{d}x^{\alpha}}{\mathrm{d}s}\frac{\mathrm{d}x^{\beta}}{\mathrm{d}s} = 0, \label{geodesicfull}
\fe
with \( s \) being an arbitrary parameter. The primary objective of this analysis is to explore how the quantum gravity modification parameter \(\xi\) affects the motion of massless particles. Achieving this involves tackling a system of complex partial differential equations derived from Eq. (\ref{geodesicfull}). Specifically, the above equation produces four interdependent partial differential equations that need to be solved. 
Considering Eq. \ref{metric} - \ref{model2} these equations for both models are as follow\\
Model  \Romannum{1}:\\
\ie
\frac{\mathrm{d}t^{\prime}}{\mathrm{d}s} =
-\frac{r' t' \left(r^3 f'(r)+2 \xi ^2 r f(r) f'(r)-2 \xi ^2 f(r)^2\right)}{r f(r) \left(\xi ^2 f(r)+r^2\right)}, 
\fe
\begin{equation}
\begin{split}
\frac{\mathrm{d}r^{\prime}}{\mathrm{d}s}  = &\frac{\left(r'\right)^2 \left(r^3 f'(r)+2 \xi ^2 r f(r) f'(r)-2 \xi ^2 f(r)^2\right)}{2 r f(r) \left(\xi ^2 f(r)+r^2\right)}+\\
&r \left(\theta '\right)^2 \left(\frac{\xi ^2 f(r)^2}{r^2}+f(r)\right)+\frac{f(r) \sin ^2(\theta ) \left(\varphi '\right)^2 \left(\xi ^2 f(r)+r^2\right)}{r},
\end{split}
\end{equation}
Model \Romannum{2}:\\
\ie
\frac{\mathrm{d}t^{\prime}}{\mathrm{d}s} = -\frac{r' t' f'(r)}{f(r)},
\fe
\begin{equation}
\begin{split}
\frac{\mathrm{d}r^{\prime}}{\mathrm{d}s}  = &  -\frac{1}{2} f(r) \left(t'\right)^2 \left(\frac{\xi ^2 f(r)}{r^2}+1\right) f'(r)+\frac{\left(r'\right)^2 \left(r^3 f'(r)+2 \xi ^2 r f(r) f'(r)-2 \xi ^2 f(r)^2\right)}{2 r f(r) \left(\xi ^2 f(r)+r^2\right)}+\\
&r f(r) \left(\theta '\right)^2 \left(\frac{\xi ^2 f(r)}{r^2}+1\right)+r f(r) \sin ^2(\theta ) \left(\varphi '\right)^2 \left(\frac{\xi ^2 f(r)}{r^2}+1\right),
\end{split}
\end{equation}
and for both models:
\ie
\frac{\mathrm{d}\theta^{\prime}}{\mathrm{d}s} = \sin (\theta ) \cos (\theta ) \left(\varphi '\right)^2-\frac{2 \theta ' r'}{r},
\fe
and
\ie
\frac{\mathrm{d}\varphi^{\prime}}{\mathrm{d}s} = -\frac{2 \varphi ' \left(r'+r \theta ' \cot (\theta )\right)}{r}.
\fe
Here, the prime symbol (\('\)) denotes differentiation with respect to \(s\) (i.e., \(\mathrm{d}/\mathrm{d}s\)) and $f(r) = 1 - \frac{2M}{r}$.
%Fig. \ref{geodesicsmodeli} displays the geodesic trajectories calculated numerically for \(M = 1\) and several values of \(\xi\). The orange dashed lines correspond to the photon sphere, while the light paths for varying \(\xi\) are represented by distinct solid lines in different colors.

\begin{figure}
    \centering
     \includegraphics[width=75mm]{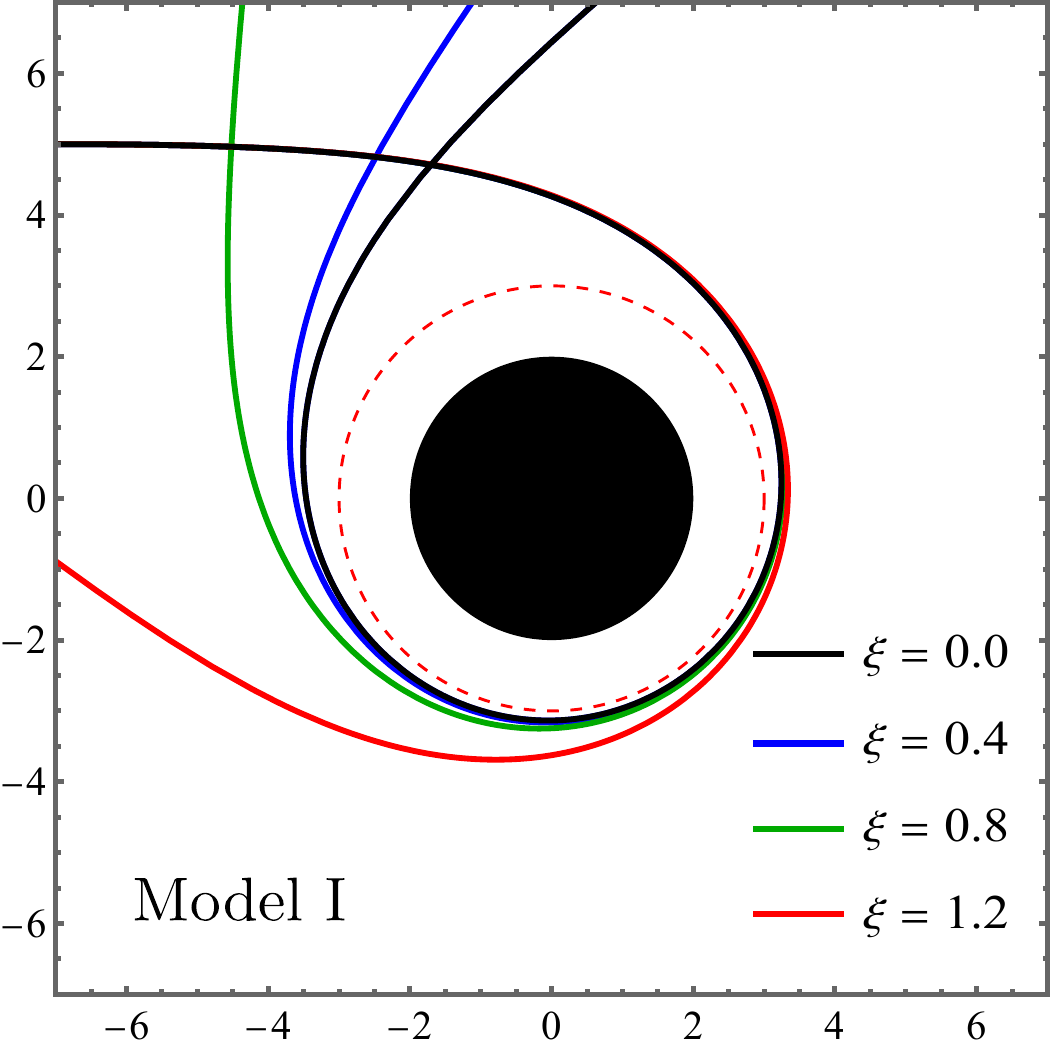}
     \includegraphics[width=75mm]{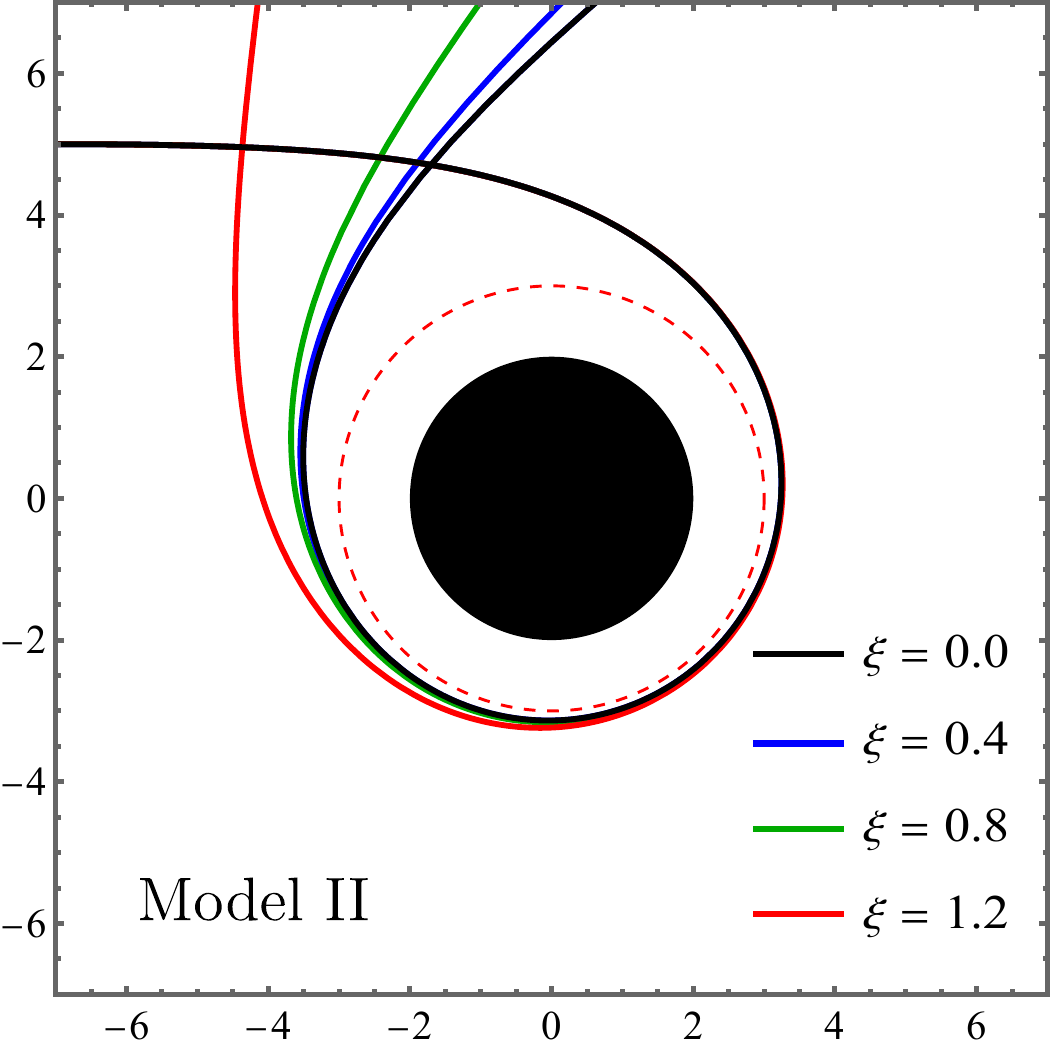}
    \caption{The geodesic trajectories are numerically calculated using \(M = 1\), b = 5 and different values of \(\xi\) for model \Romannum{1}. The orange dashed circle represents the photonic radius.}
    \label{geodesicsmodeli}
\end{figure}

%%%%%%%%%%%%%%%%%%%%%%%%%%%%%%%%%%%%%%%%%%%%%%%%%%%%%%%%%%%%%%%%%%%%%%%%%%%%%%%%%%%%%%%%%%%%%%%%%%%%%%%%%%%%%%%%%%%%%%%%%%%%%%%%%%%%%%%%%%%%%%%%%%%%%%%%%%%%%%%%%%%%%%%%%%%%%%%%%%%%%%%%%%%%%%%%%%%%%%%%%%%%%%%%%%%%%%%%%%%%%%%%%%%%%%%%%%%%%%%%%%%%%%

%\subsection{Model II}

%Analogously to the model I, we have also four differential equations to solve numerically. These are given below:
%\ie
%\frac{\mathrm{d}t^{\prime}}{\mathrm{d}s} = -\frac{r' t' f'(r)}{f(r)},
%\fe
%\begin{equation}
%\begin{split}
%\frac{\mathrm{d}r^{\prime}}{\mathrm{d}s}  = &  -\frac{1}{2} f(r) \left(t'\right)^2 \left(\frac{\xi ^2 f(r)}{r^2}+1\right) f'(r)+\frac{\left(r'\right)^2 \left(r^3 f'(r)+2 \xi ^2 r f(r) f'(r)-2 \xi ^2 f(r)^2\right)}{2 r f(r) \left(\xi ^2 f(r)+r^2\right)}+\\
%&r f(r) \left(\theta '\right)^2 \left(\frac{\xi ^2 f(r)}{r^2}+1\right)+r f(r) \sin ^2(\theta ) \left(\varphi '\right)^2 \left(\frac{\xi ^2 f(r)}{r^2}+1\right),
%\end{split}
%\end{equation}
%\ie
%\frac{\mathrm{d}\theta^{\prime}}{\mathrm{d}s} = \sin (\theta ) \cos (\theta ) \left(\varphi '\right)^2-\frac{2 \theta ' r'}{r},
%\fe
%and
%\ie
%\frac{\mathrm{d}\varphi^{\prime}}{\mathrm{d}s} = -\frac{2 \varphi ' \left(r'+r \theta ' \cot (\theta )\right)}{r}.
%\fe

Fig. \ref{geodesicsmodeli} shows the numerically calculated geodesic paths in model I and model II, for \(M =1\) and $\xi = 0$, $0.4$, $0.8$ and $1.2$. The orange dashed lines indicate the photon sphere. The colored solid line indicates the path of light.

As shown in Figs. \ref{geodesicsmodeli}, light trajectories are significantly influenced by the effective quantum gravity framework in both models. Additionally, the nature of this influence differs notably between models I and II. For example, when comparing the light path for $\xi = 1.2$ (indicated in red) in the left and right panel of Fig. \ref{geodesicsmodeli}, it becomes evident that the gravitational effects on the light in model II are more pronounced. This suggests that the quantum parameter $\xi$ plays a crucial role in altering the gravitational lensing effect in these models, a topic that will be further explored in the following section.

%%%%%%%%%%%%%%%%%%%%%%%%%%%%%%%%%%%%%%%%%%%%%%%%%%%%%%%%%%%%%%%%%%%%%%%%%%%%%%%%%%%%%%%%%%%%%%%%%%%%%%%%

\section{Photonspheres and shadows - an analytical consideration}\label{Shadow}
According to the spacetime elements in Eq. \ref{metric}, the photon sphere can be found using the equation \cite{Claudel:2000yi,Virbhadra:2002ju}
\begin{equation}\label{Eqrph}
    A'(r)C(r) - A(r)C'(r) = 0.
\end{equation}
Furthermore, due to the presence of two Killing vectors, the system has two conserved quantities: the energy, given by \( E = A(r) \Dot{t} \), and the angular momentum, expressed as \( L = C(r) \Dot{\phi} \). It is assumed that both \(E\) and \(L\) are nonzero. Using these quantities, we define the impact parameter \( b \) as follows
\ie \label{impact}
b \equiv \frac{L}{E} = \frac{C(r)\Dot{\phi}}{A(r)\Dot{t}},
\fe
which has the following form for the photonic radius called critical impact parameter 
\ie \label{bcrit0}
b_{\rm{c}}= \sqrt{\frac{C{ (r_{\rm ph}})}{A (r_{\rm ph})}}.
\fe
Now, in the second model, Eq. \eqref{Eqrph} reduces to \( r_{\rm ph} = 3M \). In contrast, the first model yields four solutions, among which two are physically relevant: \( r_{\rm ph} = 3M \), corresponding to the standard Schwarzschild result, which remains unchanged, and
\begin{equation} \label{rph}
    r_{\rm ph} = \frac{(2\eta)^{1/3}}{3} - \frac{2\xi^2}{(2\eta)^{1/3}}.
\end{equation}
Following the same approach used in the analytical analysis of the horizon, we investigate Eq. \eqref{rph} under specific limits. When \(\xi \rightarrow 0\), the photon sphere radius tends to zero, while the Schwarzschild case, \( r_{\rm ph} = 3M \), persists. In the limit \(\xi \rightarrow \infty\), we identify that the second photon sphere is given by
\begin{equation}
    r_{\rm ph} = 2M - \frac{4M^3}{\xi^2}.
\end{equation}
Observe that this second photon sphere coincides with the event horizon at \( 2M \). Consequently, the corresponding impact parameter is negative, indicating that this solution for \( r_{\rm ph} \) does not contribute to shadow formation.
Since the function \( A(r) \) remains unchanged from the Schwarzschild case in model \Romannum{2}, the photon sphere radius and critical impact parameter are preserved as \( r_{\rm ph} = 3M \) and \( b_ \rm c = 3\sqrt{3}M \), respectively. In contrast, in model I, while the temporal component of the spacetime depends on \(\xi\), the photon sphere radius is still \(3M\), similar to model II. However, the impact parameter becomes \(\xi\)-dependent and is given by
\begin{equation}\label{bcrit}
    b_{\rm{c}}^2 = \frac{729M^4}{27M^2+\xi^2}.
\end{equation}
Then, the exact expression for the observer--dependent shadow radius is
\begin{align}
    R_{\rm sh} &= \sqrt{b_{\rm{c}}^2 A(r_{\rm obs})} \nonumber \\
    &= \sqrt{\frac{729M^4}{27M^2+\xi^2}\left\{ \left(1 - \frac{2M}{r_{\rm obs}} \right)\left[1+\frac{\xi^2}{r^2_{\rm obs}}\left( 1 - \frac{2M}{r_{\rm obs}}\right) \right] \right\}}.
\end{align}
We aim to reexamine and constrain the parameter \(\xi\) using the Event Horizon Telescope (EHT) observations for Sgr. A$^*$ and M87$^*$. To this end, it is essential to approximate the above equation in the limit \( r_{\rm obs} \rightarrow \infty \). The resulting expression is given by
\begin{equation}
    R_{\rm sh} \sim \frac{27 M^{2}}{\sqrt{27 M^{2}+\xi^{2}}}-\frac{27 M^{3}}{\sqrt{27 M^{2}+\xi^{2}}\, r_{\rm obs}} + \mathcal{O}(r^{-2}_{\rm obs}).
\end{equation}
Note that the first term is similar to the one that is found in Ref. \cite{Konoplya:2024lch} but without the consideration of the second term correction. To study the behavior of the shadow analytically, we again perform some approximations for $\xi$. When $\xi \rightarrow 0$, the shadow radius is given approximately as
\begin{equation}
    R_{\rm sh} \sim 3 M \sqrt{3}-\frac{3 M^{2} \sqrt{3}}{r_{\rm obs}}+\left(\frac{1}{r_{\rm obs}}-\frac{1}{M}\right)\frac{ \xi^{2} \sqrt{3}}{18} + \mathcal{O}(\xi^4).
\end{equation}
Such a result is interesting since in the weak field regime, the shadow is altered by a small value of $\xi$, as indicated by the third term in the above equation. This is in contrast to the strong field regime phenomena such as the horizon and the photonsphere, since if $\xi$ is small, the classical case remains. Now, when $\xi$ is large, we have seen deviations from the classical case for the horizon and photonsphere. But for the shadow, however,
\begin{equation}
    R_{\rm sh} \sim \frac{27 M^{2}}{\xi}-\frac{27 M^{3}}{\xi  r_{\rm obs}} - \mathcal{O}(\xi^{-3}),
\end{equation}
confirming that the shadow formation vanishes. That is, one cannot detect the influence of large values of $\xi$ in the weak field regime as far as the shadow formation is concerned.
The Schwarzschild shadow radius is bounded by the following uncertainties: $4.209M \leq R_{\rm sh} \leq 5.560M$ for Sgr. A$^*$ at $2\sigma$ level of significance \cite{Vagnozzi:2022moj}, and $ 4.313M \leq R_{\rm sh} \leq 6.079M$ for M87$^*$ at $1\sigma$ level of significance \cite{EventHorizonTelescope:2021dqv}. Let $\delta$ represent the difference between upper (or lower) bounds and $R_{\rm sh}$. Then we can calculate the constraint in the parameter $\xi$ as
\begin{equation} \label{Rsh_cons}
    \xi = \pm \sqrt{-6\sqrt{3}\delta M}.
\end{equation}
Results implied that there are no upper bounds for $\xi$, albeit negative values are also allowed. For instance, for Sgr. A$^*$, $\delta = -0.987$ giving $\xi=\pm3.203M$. For M87$^*$, $\delta = -0.883$ leads to $\xi = \pm3.030M$.

%%%%%%%%%%%%%%%%%%%%%%%%%%%%%%%%%%%%%%%%%%%%%%%%%%%%%%%%%%%%%%%%%%%%%%%%%%%%%%%%%%%%%%%%%%%%%%%%%%%%%%%%%%%%%%%%%%%%%%%%%%%%%%%%%%%%%%%%%%%%%%%%%%%%%%%%%%%%%%%%%%%%%%%%%%%%%%%%%%%%%%%%%%%%%%%%

\section{Gravitational lensing}

Recently, studies have appeared in the literature employing the Bozza method to investigate gravitational lensing in the strong deflection regime \cite{Li:2024afr,liu2024gravitationals,Li:2024afr} for the black hole under consideration \cite{zhang2024black}. Complementing these approaches, our work explores the lensing phenomena in the weak deflection regime using the \textit{Gauss--Bonnet} theorem \cite{Gibbons:2008rj}. Furthermore, we extend our analysis to the strong deflection regime by utilizing the Tsukamoto technique \cite{tsukamoto2017deflection}. 
To better organize our results, we shall present the calculations separately for model I and model II in the following subsections.

%%%%%%%%%%%%%%%%%%%%%%%%%%%%%%%%%%%%%%%%%%%%%%%%%%%%%%%%%%%%%%%%%%%%%%%%%%%%%%%%%%%%%%%%%%%%%%%%%%%%%%%%%%%%%%%%%%%%%%%%%%%%%%%%%%%%%%%%%%%%%%%%%%%%%%%%%%%%%%%%%%%%%%%%%%%%%%%%%%%%%%%%%%%%%%%%%%%

\subsection{Weak deflection limit}

\subsubsection{Model I}

\begin{figure}
    \centering
     \includegraphics[width=0.55\textwidth]{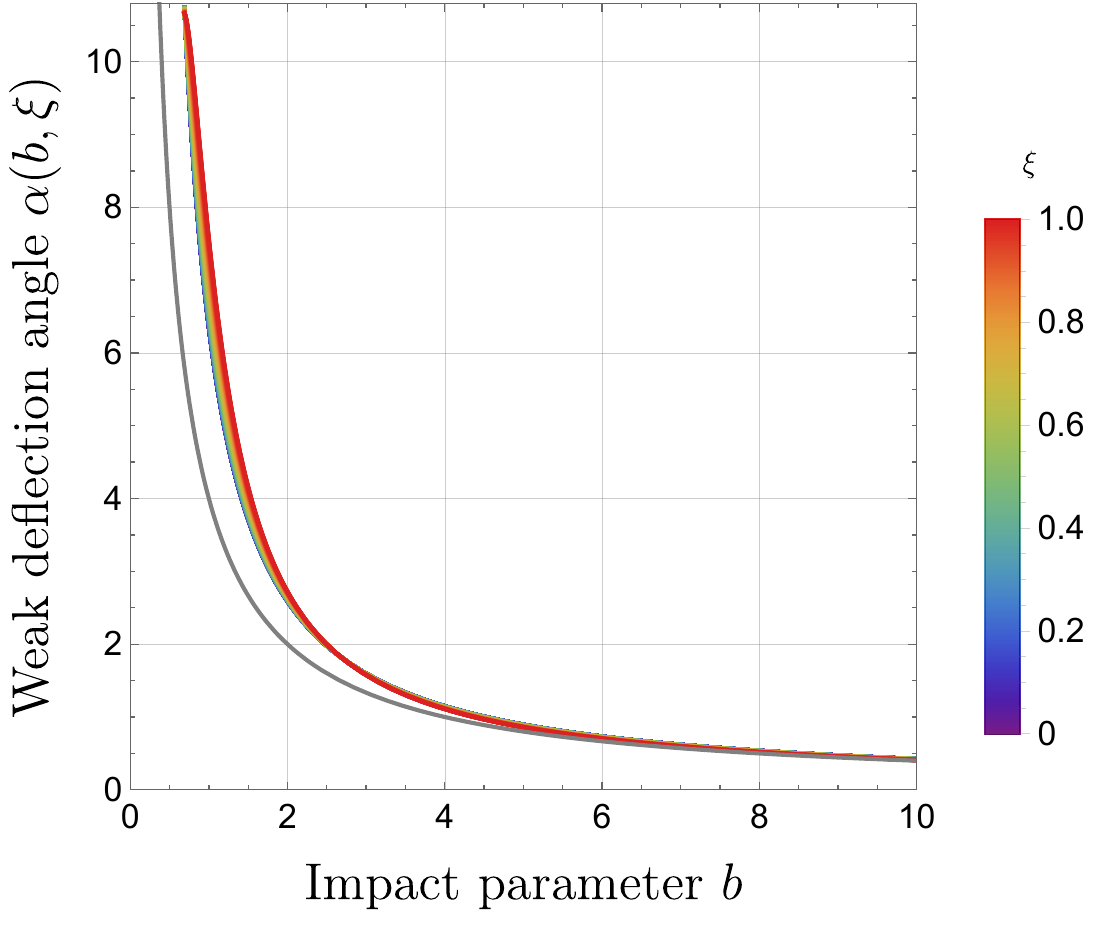}
    \caption{The deflection angle as a function of $b$ for different values of $\xi$ (considering a fixed value of $M=1$). The Schwarzschild case is represented by the gray line.}
    \label{gbwdl}
\end{figure}

This section revisits the \textit{Gauss--Bonnet }theorem and applies it to determine the black hole's weak deflection angle. To begin, we consider the null geodesic condition, \(\mathrm{d}s^2 = 0\), and manipulate it to obtain the following expression
\begin{eqnarray}
\mathrm{d}t^2=\gamma_{ij}\mathrm{d}x^i \mathrm{d}x^j=\frac{1}{A(r)^2}\mathrm{d}r^2+\frac{r^2}{A(r)}\mathrm{d}\Omega^2,~\label{opmetric}
\end{eqnarray}
with the indices \(i\) and \(j\) span from $1$ to $3$, and \(\gamma_{ij}\) denotes the components of the optical metric. To apply the \textit{Gauss--Bonnet} theorem, it becomes essential to evaluate the Gaussian curvature, which is determined as follows
\begin{equation}
\begin{split}
\mathcal{K}=\frac{R}{2} & = \frac{A(r)}{2} \frac{\mathrm{d}^{2}}{\mathrm{d} r^{2}}A(r) -\frac{\left(\frac{\mathrm{d}}{\mathrm{d} r}A(r)\right)^{2}}{4} \\
 & =   \frac{6 M^4 \xi ^4}{r^{10}}-\frac{20 M^3 \xi ^4}{r^9}+\frac{24 M^2 \xi ^4}{r^8}-\frac{9 M^3 \xi ^2}{r^7}-\frac{12 M \xi ^4}{r^7}\\ 
& +\frac{21 M^2 \xi ^2}{r^6} 
 +\frac{2 \xi ^4}{r^6}-\frac{15 M \xi ^2}{r^5}+\frac{3 M^2}{4 r^4}+\frac{3 \xi ^2}{r^4}-\frac{M}{r^3}.
\end{split}
\end{equation}
In this context, \(\gamma\) stands for the determinant of the optical metric components \(\gamma_{ij}\), while \(R\) denotes the Ricci scalar. The surface area confined to the equatorial plane is given by \cite{Gibbons:2008rj}:
\ie
\mathrm{d}S=\sqrt\gamma \mathrm{d}r \mathrm{d} \phi= \frac{r}{A(r)^{3/2}} \mathrm{d}r \mathrm{d}\phi = \left[\frac{r}{\left(\left(1-\frac{2 M}{r}\right) \left(\frac{\xi ^2 \left(1-\frac{2 M}{r}\right)}{r^2}+1\right)\right)^{3/2}}\right] \mathrm{d}r \mathrm{d}\phi.
\label{dsss}
\fe
The resulting deflection angle of light is then determined by the following expression:
\begin{eqnarray}
\alpha&=&-\int\int_{\tilde{D}}\mathcal{K}\mathrm{d}S=-\int^{\pi}_0\int^{\infty}_{\frac{b}{\sin\phi}}\mathcal{K}\mathrm{d}S \nonumber\\
&\simeq&  \frac{4 M}{b} + \frac{3 \pi  M^2}{4 b^2} +\frac{8 M \xi ^2}{b^3} -\frac{3 \pi  \xi ^2}{4 b^2} -\frac{45 \pi  M^2 \xi ^2}{32 b^4}.~\label{deflang}
\end{eqnarray}
It should be noted that when the parameter controlling the quantum gravity corrections tends to zero (\(\xi \to 0\)), the deflection angle converges to the classical Schwarzschild result. Fig. \ref{gbwdl} illustrates the variation of the weak deflection angle as a function of \( b \) for different values of \(\xi\), with the gray line indicating the Schwarzschild scenario for reference.

The expression in Eq. \eqref{deflang} can be further generalized to account for the finite distances of both the source and observer, as well as for the deflection angle of massive particles. This extension is accomplished using the approach proposed by Li et al. \cite{Li:2020wvn}, which adapts the \textit{Gauss--Bonnet} theorem to scenarios involving non--asymptotically flat spacetimes. Consequently, the weak deflection angle \((\alpha_{\rm wda})\) is obtained by solving:
\begin{equation} \label{wda_Li}
    \alpha_{\rm wda} = \iint_{_{r_{\rm ph}}^{\rm R }\square _{r_{\rm ph}}^{\rm S}}K\mathrm{d}S + \phi_{\text{RS}} = \int^{\phi_{\rm R}}_{\phi_{\rm S}} \int_{r_{\rm ph}}^{r(\phi)} K\sqrt{g} \, \mathrm{d}r \, \mathrm{d}\phi + \phi_{\rm RS}.
\end{equation}
The term \(\phi_{\rm RS} = \phi_{\rm R} - \phi_{\rm S}\) represents the separation angle in the equatorial plane between the source and the receiver, where \(\phi_{\rm R} = \pi - \phi_{\rm S}\). Here, \(K\) denotes the Gaussian curvature, expressed as
\begin{equation}
    K=\frac{1}{\sqrt{g}}\left[\frac{\partial}{\partial\phi}\left(\frac{\sqrt{g}}{g_{rr}}\Gamma_{rr}^{\phi}\right)-\frac{\partial}{\partial r}\left(\frac{\sqrt{g}}{g_{rr}}\Gamma_{r\phi}^{\phi}\right)\right] 
    =-\frac{1}{\sqrt{g}}\left[\frac{\partial}{\partial r}\left(\frac{\sqrt{g}}{g_{rr}}\Gamma_{r\phi}^{\phi}\right)\right].
\end{equation}
Given the Jacobi metric
\begin{equation} \label{Jac_met}
\mathrm{d}l^2=g_{ij}\mathrm{d}x^{i}\mathrm{d}x^{j}
    =(E^2-\mu^2A(r))\left(\frac{B(r)}{A(r)}\mathrm{d}r^2+\frac{C(r)}{A(r)}\mathrm{d}\Phi^2\right),
\end{equation}
one can find its determinant $g$ as
\begin{equation}
    g = \frac{(E^2 - \mu^2 A(r))B(r)C(r)}{A(r)^2}.
\end{equation}
The lower limit of the integral in Eq. \eqref{wda_Li} becomes zero when \( r = r_{\rm ph} \). Therefore, we have:
\begin{equation} \label{gct}
    \int_{r_{\rm ph}}^{r(\phi)} K\sqrt{g}\mathrm{d}r = -\frac{A(r)\left(E^{2}-A(r)\right)C'-E^{2}C(r)A(r)'}{2A(r)\left(E^{2}-A(r)\right)\sqrt{B(r)C(r)}}\bigg|_{r = r(\phi)},
\end{equation}
where $E$ is defined as the energy per unit mass of the particle.

To evaluate Eq. \eqref{gct}, we need the orbit equation to determine the point of closest approach to the black hole. Following standard procedures in celestial mechanics, we define \( u = r^{-1} \). Thus, we obtain:
\begin{equation}
    F(u) \equiv \left(\frac{\mathrm{d}u}{\mathrm{d}\varphi}\right)^2 
    = \frac{C(u)^2u^4}{A(u)B(u)}\Bigg[\left(\frac{E}{J}\right)^2-A(u)\left(\frac{1}{J^2}+\frac{1}{C(u)}\right)\Bigg],
\end{equation}
where \( J = Evb \) represents the angular momentum per unit mass of the particle, with \( b \) denoting the impact parameter. The conditions \(\frac{\mathrm{d}u}{\mathrm{d}\phi} = \frac{\mathrm{d}^2u}{\mathrm{d}\phi^2} = 0\) must be satisfied to ensure a stable orbit. Solving the resulting differential equation under these constraints yields:
\begin{equation}
    u(\phi) = \frac{1}{b}\sin \! \left(\phi \right)+ \frac{M}{b^{2} v^{2}}\left[1 + v^2 \cos \! \left(\phi \right)\right].
\end{equation}
To incorporate the parameter \(\xi\) into \(u(\phi)\), we employ an iterative solution method, which results in the following expression:
\begin{equation} \label{itera_u}
    u(\phi) = \frac{1}{b}\sin \! \left(\phi \right)+ \frac{M}{b^{2} v^{2}}\left[1 + v^2 \cos \! \left(\phi \right)\right] - \frac{\xi^2}{2v^2b^3}.
\end{equation}
Substituting this equation into Eq. \eqref{gct}, we obtain
\begin{equation}
    \int_{r_{\rm ph}}^{r(\phi)} K\sqrt{g}\mathrm{d}r \sim \frac{\left(2 E^{2}-1\right) M}{\left(E^{2}-1\right) b} \sin \! \left(\phi\right) - 1 - {\frac { \left( 3\,{E}^{2}-1 \right) {\xi}^{2}}{{b}^{2} \left( 2\,{E}^{2}-2 \right) }} \sin \! \left(\phi\right)^2 + \mathcal{O}(M\xi^2).
\end{equation}
By integrating the above expression with respect to \(\mathrm{d}\phi\), we get
\begin{align}
    \int_{\phi_{\rm S}}^{\phi_{\rm R}} \int_{r_{\rm ph}}^{r(\phi)} K\sqrt{g} \, \mathrm{d}r \, \mathrm{d}\phi & \sim -\frac{\left(2 E^{2}-1\right) M}{\left(E^{2}-1\right) b}\cos \! \left(\phi\right) \bigg\vert_{\phi_{\rm S}}^{\phi_{\rm R}} -\phi_{\rm RS} \nonumber \\
    &- {\frac { \left( 3\,{E}^{2}-1 \right) {\xi}^{2}}{4{b}^{2} \left( \,{E}^{2}-2 \right) }} \left( \phi_{\rm RS} - \sin(\phi)\cos(\phi)\bigg\vert_{\phi_{\rm S}}^{\phi_{\rm R}}  \right) + \mathcal{O}(M\xi^2).
\end{align}
It is important to point out that solving for \(\phi\) is necessary in the above context. By proceeding with this, and utilizing Eq.\eqref{itera_u},
\begin{equation}
    \phi = \arcsin(bu)+\frac{M\left[v^{2}\left(b^{2}u^{2}-1\right)-1\right]}{bv^{2}\sqrt{1-b^{2}u^{2}}} + \frac{\xi^{2}}{2 b^{2} v^{2} \sqrt{1-b^{2} u^{2}}} + \mathcal{O}(M\xi^2).
\end{equation}
For simplicity, we now assume that both the source and the receiver are equidistant from the black hole. Thus, we can set \(\phi_{\rm S} = \phi\). Under this assumption, we obtain the following expression:
\begin{equation}
    \cos(\phi) = \sqrt{1-b^{2}u^{2}}-\frac{M u\left[v^{2}\left(b^{2}u^{2}-1\right)-1\right]}{\sqrt{v^{2}\left(1-b^{2}u^{2}\right)}} - \frac{u \,\xi^{2}}{2 b \,v^{2} \sqrt{1-b^{2} u^{2}}} - \mathcal{O}(M\xi^2),
\end{equation}
and
\begin{equation}
    \cos(\phi) \sin(\phi) = \sqrt{1-b^{2} u^{2}}\, b u - \frac{\left(2 b^{2} u^{2}-1\right) \left(b^{2} u^{2} v^{2}-v^{2}-1\right) M}{\sqrt{1-b^{2} u^{2}}\, b \,v^{2}} + \frac{\left(1-2 b^{2} u^{2}\right) \xi^{2}}{2 b^{2} v^{2} \sqrt{1-b^{2} u^{2}}} - \mathcal{O}(M\xi^2).
\end{equation}
By substituting Eq. \eqref{wda_Li} along with the two preceding equations, we derive the general formula for the weak deflection angle in model I as 
\begin{equation}
    \alpha_{\rm wda} \sim \frac{2 M\left(v^{2}+1\right) \sqrt{1-b^{2} u^{2}}}{v^{2} b} - \frac{\left(v^{2}+2\right) \xi^{2}}{4 b^{2} v^{2}} \left[ 2 \sqrt{1-b^{2} u^{2}}\, b u +\pi -2 \arcsin \! \left(b u \right) \right] + \mathcal{O}(M\xi^2).
\end{equation}
In the far-field approximation, where \( u \rightarrow 0 \), the above expression simplifies to:
\begin{equation}
    \alpha_{\rm wda} \sim \frac{2 M \left(v^{2}+1\right)}{v^{2} b}-\frac{\left(v^{2}+2\right) \xi^{2} \pi}{4 b^{2} v^{2}} + \mathcal{O}(M\xi^2).
\end{equation}
Finally, for photons, where \( v = 1 \), the expression becomes
\begin{equation}
    \alpha_{\rm} \sim \frac{4M}{b} - \frac{3 \xi^{2} \pi}{4 b^{2}} + \mathcal{O}(M\xi^2).
\end{equation}
It is noteworthy that the final expression aligns with Eq. \eqref{deflang} when considering only the linear contributions, as these terms represent the dominant effects.

%%%%%%%%%%%%%%%%%%%%%%%%%%%%%%%%%%%%%%%%%%%%%%%%%%%%%%%%%%%%%%%%%%%%%%%%%%%%%%%%%%%%%%%%%%%%%%%%%%%%%%%%%%%%%%%%%%%%%%%%%%%%%%%%%%%%%%%%%%%%%%%%%%%%%%%%%%%%%%%%%%%%%%%%%%%%%%%%%%%%%%%%%%%%%%%%%%%%%%%%%%%%%%%%%%%%%%%%%%%%%%%%%%%%%%%%%%%%%%%%%%%%%%%%%%%%%%%%%%%%%%%%%%%%%%%%%%%%%%%%%%%%%%%%%%%%%%%%%%%%%%%%%%%%%%%%%%%%%%%%%%%%%%%%%%%%%%%%%%%%%%%%%%%%%%%%%%%%%%%%%%%%%%%%%%%%%%%%%%%%%%%%%%%%%%%%%%%%

\subsubsection{Model II}

For the alternative model examined in this paper, the procedure to derive the weak deflection angle follows the same steps as for model I. The resulting expression is
\begin{equation}
    \alpha_{\rm wda} \sim \frac{2 M\left(v^{2}+1\right) \sqrt{1-b^{2} u^{2}}}{v^{2} b} - \frac{\xi^{2} \left(2 \sqrt{1-b^{2} u^{2}}\, b u +\pi -2 \arcsin \! \left(b u \right)\right)}{4 b^{2}} + \mathcal{O}(M\xi^2).
\end{equation}
Here, we observe that the second term does not contain \(v\). Thus, in the far--field approximation, it simplifies to:
\begin{equation}
    \alpha_{\rm wda} \sim \frac{2 \left(v^{2}+1\right) M}{v^{2} b}-\frac{\xi^{2} \pi}{4 b^{2}} + \mathcal{O}(M\xi^2).
\end{equation}
When $v = 1$,
\begin{equation}
    \alpha_{\rm wda} \sim \frac{4M}{b} -\frac{\xi^{2} \pi}{4 b^{2}} + \mathcal{O}(M\xi^2).
\end{equation}
The key distinction between this result and that of model I is the presence of a factor of \(3\) in the second term.

%%%%%%%%%%%%%%%%%%%%%%%%%%%%%%%%%%%%%%%%%%%%%%%%%%%%%%%%%%%%%%%%%%%%%%%%%%%%%%%%%%%%%%%%%%%%%%%%%%%%%%%%%%%%%%%%%%%%%%%%%%%%%%%%%%%%%%%%%%%%%%%%%%%%%%%%%%%%%%%%%%%%%%%%%%%%%%%%%%%%%%%%%%%%%%%%%%%%%%%%%%%%%%%%%%%%%%%%%%%%%%%%%%%%%%%%%%%%%%%%%%%%%%%%%%%%%%%%%%%%%%%%%%%%%%%%%%%%%%%%%%%%%%%%%%%%%%%%%%%%%%%%%%%%%%%%%%%%%%%%%%%%%%%%%%%%%%%%%%%%%%%%%%%%%%%%%%%%%%%%%%%%%%%%%%%%%%%%%%%%%%%%%%%%%%%%%%%%%%%%%%%%%%%%%%%%%%%%%%%%%%%%%%%%%%%%%%%%%%%%%%%%%%%%%%%%%%%%%%%%%%%%%%%%%%%%%%%%%%%%%%%%

\subsection{Strong deflection limit}

This section outlines the general approach employed to compute the deflection angle of a light ray under the strong deflection limit \cite{tsukamoto2017deflection}. As in previous studies, the analysis is centered on asymptotically flat, static, and spherically symmetric spacetimes, which are described by the line element in Eq. \ref{metric}.
To utilize the method introduced by Tsukamoto \cite{tsukamoto2017deflection}, the metric must meet the requirement of asymptotic flatness as \( A(r) \), \( B(r) \), and \( C(r) \) exhibit the following asymptotic behavior: \(\lim\limits_{r \to \infty}A(r) = 1\), \(\lim\limits_{r \to \infty}B(r) = 1\), and \(\lim\limits_{r \to \infty}C(r) = r^{2}\) which are fulfilled in the effective quantum gravity models. Additionally, due to the inherent symmetries of the spacetime, there exist two associated Killing vectors: \(\partial_t\) and \(\partial_\phi\). 

We now outline the process for determining the deflection angle in the strong field regime, beginning with the introduction of a new variable, denoted as \( D(r) \):
\ie
D(r) \equiv \frac{C^{\prime}(r)}{C(r)} - \frac{A^{\prime}(r)}{A(r)}.
\fe
Here, it is important to mention that the prime symbols indicate differentiation with respect to the radial coordinate. It is assumed that the equation \(D(r) = 0\) has at least one positive root. The radius of the photon sphere, denoted by \(r_{\rm ph}\), corresponds to the largest positive root of \(D(r) = 0\) as discussed in Eq. \ref{rph}. We further assume that the metric functions \(A(r)\), \(B(r)\), and \(C(r)\) remain finite and positive for \(r \geq r_{\rm ph}\).

Owing to the axial symmetry of the system, the motion can be confined to the equatorial plane (\(\theta = \pi/2\)). Under this condition, the radial equation simplifies to:
\ie
\Dot{r}^{2} = V(r),
\fe
where \(V(r) = \frac{L^2 R(r)}{B(r)C(r)}\) represents the effective potential, where \(R(r)\) is defined as \( R(r) = \frac{C(r)}{A(r) b^{2}} - 1\). This equation is analogous to the motion of a particle with unit mass in a potential \(V(r)\). The photon’s trajectory is allowed wherever \(V(r) \geq 0\). Under the asymptotic flatness conditions, we have \(\lim\limits_{{r \to \infty}} V(r) = E^{2} > 0\), indicating that the photon can exist at infinity (\(r \to \infty\)). We assume that \(R(r) = 0\) has at least one positive root.

This work focuses on a gravitational lensing scenario in which a photon, originating from infinity, travels towards a gravitational source, reaches a minimum distance of \(r_0\), and then scatters back to infinity. For this scattering process to occur, the condition \(r_0 > r_{\rm ph}\) must be satisfied, since closed orbits are not possible for the photon in this configuration. The distance \(r_0\) corresponds to the largest positive root of the equation \(R(r) = 0\), where the functions \(B(r)\) and \(C(r)\) remain finite. As a result, the effective potential \(V(r)\) becomes zero at \(r = r_0\). Given that \(r_0\) is the point of closest approach, satisfying \(R(r) = 0\), it follows that:
\ie
A_{0}\Dot{t}^{2}_{0} = C_{0}\Dot{\phi}^{2}_{0}.
\fe

In this context, and for the remainder of this analysis, the subscript \( ``0" \) will denote quantities evaluated at \( r = r_0 \). To simplify the discussion, we can, without loss of generality, assume that the impact parameter \( b \) is positive, particularly when considering the path of an individual light ray, defined in Eq. \ref{bcrit0}. 
It is convenient writing $R(r)$ as
\ie
R(r)= \frac{A_{0}C}{AC_{0}} - 1.
\fe

We establish a criterion, based on the method presented in Ref. \cite{hasse2002gravitational}, that is both necessary and sufficient for a circular light orbit to exist. Under this condition, the equation governing the trajectory is presented below
\ie
\frac{BC \Dot{r}^{2}}{E^{2}} + b^{2} = \frac{C}{A},
\fe
so that
\ie
\ddot{r} + \frac{1}{2}\left( \frac{B^{\prime}}{B} + \frac{C^{\prime}}{C} \Dot{r}^{2} \right) = \frac{E^{2}D(r)}{AB}. 
\fe
For \( r \geq r_{\rm ph} \), the functions \( A(r) \), \( B(r) \), and \( C(r) \) are finite and positive, along with \( E \) also being positive. The condition \( D(r) = 0 \) guarantees the presence of a stable circular light orbit. Additionally, it is important to highlight that \( R^{\prime}_{m} = \frac{D_{m}C_{m}A_{m}}{b^{2}} = 0 \), where the subscript \( m \) indicates that the respective quantities are evaluated at \( r = r_{\rm ph} \).

This scenario will be referred to as the strong deflection limit from this point onward. Taking the derivative of \(V(r)\) with respect to \(r\) yields
\ie
V^{\prime}(r) = \frac{L^{2}}{BC} \left[ R^{\prime} + \left( \frac{C^{\prime}}{C} - \frac{B^{\prime}}{B}   \right)   R  \right].
\fe
As \(r_{0}\) approaches \(r_{m}\) within the strong deflection limit, both \(V(r_{0})\) and \(V^{\prime}(r_{0})\) approach zero. As a result, the equation governing the trajectory simplifies to the following form:
\ie
\left(  \frac{\mathrm{d}r}{\mathrm{d}\phi}     \right)^{2} = \frac{R(r)C(r)}{B(r)}.
\fe
The deflection angle, denoted as \(\alpha(r_{0})\), is defined as 
\ie
\alpha(r_{0}) = I(r_{0}) - \pi,
\fe
with, in this regard, $I(r_{0})$ reads
\ie
I(r_{0}) \equiv 2 \int^{\infty}_{r_{0}} \frac{\mathrm{d}r}{\sqrt{\frac{R(r)C(r)}{B(r)}}}.
\fe

To continue, the first step is to address the integration process. It is worth highlighting that this step is particularly complex, as emphasized in Tsukamoto's study \cite{tsukamoto2017deflection}. Furthermore, we define the following expressions, as outlined in \cite{tsukamoto2017deflection}:
\ie
z \equiv 1 - \frac{r_{0}}{r},
\fe
in a such way that this integral can be rewritten below
\ie
I(r_{0}) = \int^{1}_{0} f(z,r_{0}) \mathrm{d}z,
\fe
where 
\ie
f(z,z_{0}) \equiv \frac{2r_{0}}{\sqrt{G(z,r_{0})}}, \,\,\,\,\,\,\,\, \text{and} \,\,\,\,\,\,\,\,  G(z,r_{0}) \equiv R \frac{C}{B}(1-z)^{4}.
\fe

Additionally, when expressed in terms of \(z\), the function \(R(r)\) takes the form:
\ie
R(r) = D_{0}r_{0} z + \left[ \frac{r_{0}}{2}\left( \frac{C^{\prime\prime}_{0}}{C_{0}} - \frac{A_{0}^{\prime\prime}}{A_{0}}  \right) + \left( 1 - \frac{A_{0}^{\prime}r_{0}}{A_{0}}  \right) D_{0}  \right] r_{0} z^{2} + \mathcal{O}(z^{3})+ ...    \,\,\,\,.
\fe
By expanding \( G(z, r_{0}) \) as a series in \( z \), we obtain:
\ie
G(z,r_{0}) = \sum^{\infty}_{n=1} c_{n}(r_{0})z^{n},
\fe
with $c_{1}(r)$ and $c_{2}(r)$ being
\ie
c_{1}(r_{0}) = \frac{C_{0}D_{0}r_{0}}{B_{0}},
\fe
and
\ie
c_{2}(r_{0}) = \frac{C_{0}r_{0}}{B_{0}} \left\{ D_{0} \left[ \left( D_{0} - \frac{B^{\prime}_{0}}{B_{0}}  \right)r_{0} -3       \right] + \frac{r_{0}}{2} \left(  \frac{C^{\prime\prime}_{0}}{C_{0}} - \frac{A^{\prime\prime}_{0}}{A_{0}}  \right)                 \right\}.
\fe

Furthermore, when applying the strong deflection limit, the expression simplifies to
\ie
c_{1}(r_{m}) = 0, \,\,\,\,\,\, \text{and} \,\,\,\,\,\, c_{2}(r_{m}) =  \frac{C_{m}r^{2}_{m}}{2 B_{m}}D^{\prime}_{m}, \,\,\,\,\,\,\, \text{with} \,\,\,\,\, D^{\prime}_{m} = \frac{C^{\prime\prime}}{C_{m}} - \frac{A^{\prime\prime}}{A_{m}}.
\fe
In shorter notation, $G(z,r_{0})$ turns out to be:
\ie
G_{m}(z) = c_{2}(r_{m})z^{2} + \mathcal{O}(z^{3}).
\fe

This indicates that the primary divergence of \(f(z, r_0)\) occurs at the \( z^{-1} \) order, leading to a logarithmic divergence in the integral \(I(r_0)\) as \(r_0\) approaches \(r_{\rm ph}\). To handle this divergence, the integral \(I(r_0)\) is separated into two components: a divergent term, \(I_D(r_0)\), and a regular term, \(I_R(r_0)\). Accordingly, the divergent portion \(I_D(r_0)\) is expressed as
\ie
I_{D}(r_{0}) \equiv \int^{1}_{0} f_{D}(z,r_{0}) \mathrm{d}z, \,\,\,\,\,\,\, \text{with} \,\,\,\,\,\,f_{D}(z,r_{0}) \equiv \frac{2 r_{0}}{\sqrt{c_{1}(r_{0})z + c_{2}(r_{0})z^{2}}}.
\fe
After performing the integration, it reads
\ie
I_{D} (r_{0}) = \frac{4 r_{0}}{\sqrt{c_{2}(r_{0})}} \ln \left[  \frac{\sqrt{c_{2}(r_{0})} + \sqrt{c_{1}(r_{0}) + c_{2}(r_{0})     }  }{\sqrt{c_{1}(r_{0})}}  \right].
\fe

Taking into account the series expansions of \(c_{1}(r_{0})\) and \(b(r_{0})\) around \(r_{0} - r_{m}\), we have:
\ie
c_{1}(r_{0}) = \frac{C_{m}r_{m}D^{\prime}_{m}}{B_{m}} (r_{0}-r_{m}) + \mathcal{O}((r_{0}-r_{m})^{2}),
\fe
and
\ie
b(r_{0}) = b_{c}(r_{m}) + \frac{1}{4} \sqrt{\frac{C_{m}}{A_{m}}}D^{\prime}_{m}(r_{0}-r_{m})^{2} + \mathcal{O}((r_{0}-r_{m})^{3}).
\fe
In other words, this leads to
\ie
\lim_{r_{0} \to r_{m}} c_{1}(r_{0})  =  \lim_{b \to b_{c}} \frac{2 C_{m} r_{m} \sqrt{D^{\prime}}}{B_{m}} \left(  \frac{b}{b_{c}} -1  \right)^{1/2}.
\fe
In this regards,, $I_{D}(b)$ results in
\ie
I_{D}(b) = - \frac{r_{m}}{\sqrt{c_{2}(r_{m})}} \ln\left[ \frac{b}{b_{c}} - 1 \right] + \frac{r_{m}}{\sqrt{c_{2}(r_{m})}}\ln \left[ r^{2}D^{\prime}_{m}\right] + \mathcal{O}[(b-b_{c})\ln(b-b_{c})].
\fe

To proceed further, we write $I_{R}(b)$ as shown below
\ie
I_{R}(b) = \int^{0}_{1} f_{R}(z,b_{c})\mathrm{d}z + \mathcal{O}[(b-b_{c})\ln(b-b_{c})].
\fe
The quantity \( f_{R} \) is defined as \( f_{R} = f(z, r_{0}) - f_{D}(z, r_{0}) \). In the context of the strong deflection limit, the deflection angle can be cast as
\ie
a(b) = - \Tilde{a} \ln \left[ \frac{b}{b_{c}}-1    \right] + \Tilde{b} + \mathcal{O}[(b-b_{c})\ln(b-b_{c})],
\label{1d2e3f4l5ections}
\fe
so that
\ie
\Tilde{a} = \sqrt{\frac{2 B_{m}A_{m}}{C^{\prime\prime}_{m}A_{m} - C_{m}A^{\prime\prime}_{m}}}, \,\,\,\,\,\,\,\, \text{and} \,\,\,\,\,\,\,\, \Tilde{b} = \Tilde{a} \ln\left[ r^{2}_{m}\left( \frac{C^{\prime\prime}}{C_{m}}  -  \frac{A^{\prime\prime}_{m}}{C_{m}} \right)   \right] + I_{R}(r_{m}) - \pi.
\fe

%%%%%%%%%%%%%%%%%%%%%%%%%%%%%%%%%%%%%%%%%%%%%%%%%%%%%%%%%%%%%%%%%%%%%%%%%%%%%%%%%%%%%%%%%%%%%%%%%%%%%%%%%%%%%%%%%%%%%%%%%%%%%%%%%%%%%%%%%%%%%%%%%%%%%%%%%%%%%%%%%%%%%%%%%%%%%%%%%%%%%%%%%%%%%%%%%%%%%%%%%%%%%%%%%%%%%%%%%%%%%%%%%%%%%%%%%%%%%%%%%%%%%%%%%%%%%%%%%%%%%%%%%%%%%%%%%%%%%%%%%%%%%%%%%%%%%%%%%%%%%%%%%%%%%%%%%%%%%%%%%%%%%%%%%%%%%%%%%%%%%%%%%%%%%%%%%%%%%%%%%%%%%%%%%%%%%%%%%%%%%%%%%%%

\subsubsection{Model I}

Applying Eq. \ref{bcrit} for model I,
\(\tilde{a}\) and \(\tilde{b}\) can be represented as
\ie
\Tilde{a} = 9 \sqrt{\frac{M^2}{81 M^2+6 \xi ^2}},
\fe
In this context, we can express it as
\ie
\begin{split}
\Tilde{b} = 9 \sqrt{\frac{M^2}{81 M^2+6 \xi ^2}} \ln \left(12-\frac{162 M^2}{27 M^2+\xi ^2}\right)
+ I_{R}(r_{m}) - \pi.
\end{split}
\fe

Unlike the situation in the Schwarzschild case, it is important to note that the contribution to the parameter \(\tilde{a}\) primarily arises from the effects of quantum gravity. Furthermore, \(I_{R}(r_{m})\) can be computed as
\ie
\begin{split}
  I_{R}(r_{m}) &  =   \int_{0}^{1} \mathrm{d}z \left\{  \frac{18 M}{\sqrt{-z^2 (2 z-3) \left(27 M^2+\xi ^2 \left((2 z-3) z^2+2\right)\right)}}-\frac{6 M}{\sqrt{z^2 \left(9 M^2+\frac{2 \xi ^2}{3}\right)}}  \right\} \\
 & = \frac{540 M^2 \ln \left(3-\sqrt{3}\right)+\xi ^2 \left(-2 \sqrt{3}+9-20 \ln (6)+20 \ln \left(\sqrt{3}+3\right)\right)}{135 M^2}.
\end{split}
\fe
It is important to highlight that an analytical result was obtained when a small value of \(\xi\) was considered.
Consequently, the deflection angle given in Eq. (\ref{1d2e3f4l5ections}) can be expressed as
\ie
\begin{split}
a(b) = &  - 9 \sqrt{\frac{M^2}{81 M^2+6 \xi ^2}} \ln \left[  \frac{b}{27 \sqrt{\frac{M^4}{27 M^2+\xi ^2}}} -1     \right] \\
& + 9 \sqrt{\frac{M^2}{81 M^2+6 \xi ^2}} \ln \left(12-\frac{162 M^2}{27 M^2+\xi ^2}\right) \\
& +\frac{540 M^2 \ln \left(3-\sqrt{3}\right)+\xi ^2 \left(-2 \sqrt{3}+9-20 \ln (6)+20 \ln \left(\sqrt{3}+3\right)\right)}{135 M^2} - \pi \\
& + \mathcal{O} \left\{ \left( b - 27 \sqrt{\frac{M^4}{27 M^2+\xi ^2}}  \right) \ln\left[\left( b - 27 \sqrt{\frac{M^4}{27 M^2+\xi ^2}}  \right) \right]  \right\}
\end{split}
\fe

\begin{figure}
    \centering
     \includegraphics[scale=0.41]{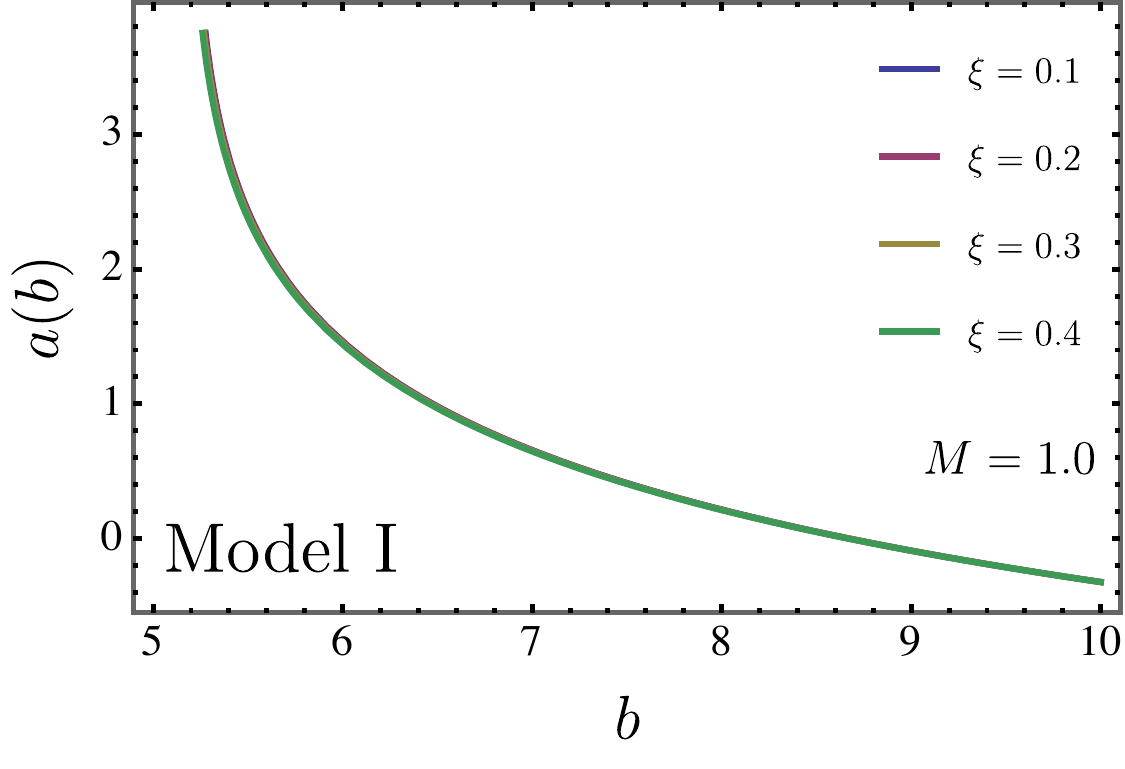}
     \includegraphics[scale=0.45]{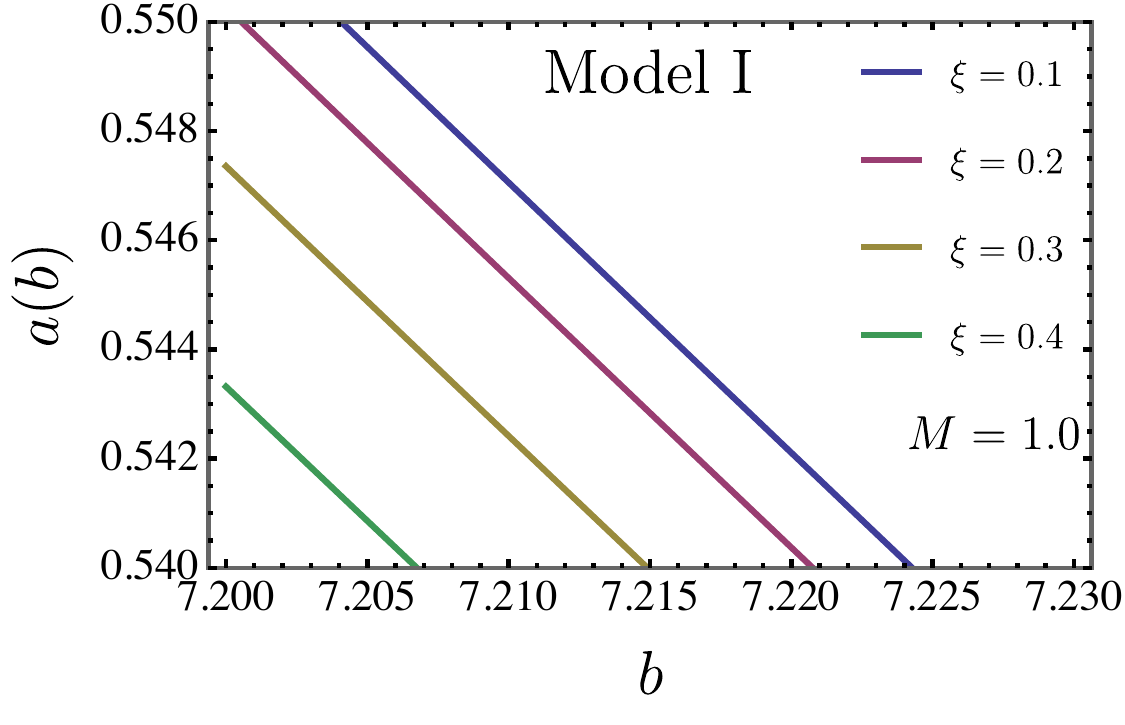}
    \caption{The strong deflection angle (Model I) as a function of $b$ for different values of $\xi$.}
    \label{strongmodeli}
\end{figure}

Fig. \ref{strongmodeli} illustrates the behavior of the strong deflection angle for model I, with \(M = 1\) held constant while the parameter \(\xi\) varies. In the left panel, the influence of \(\xi\) appears minimal, as anticipated, since these variations are indicative of quantum corrections. In contrast, the right panel clearly shows that an increase in \(\xi\) corresponds to a notable decrease in the deflection angle. Lastly, the bottom panel presents the deflection angle as a function of \(M\) for a fixed \(\xi = 0.1\), demonstrating that an increase in mass \(M\) results in a proportional increase in the deflection angle.

%%%%%%%%%%%%%%%%%%%%%%%%%%%%%%%%%%%%%%%%%%%%%%%%%%%%%%%%%%%%%%%%%%%%%%%%%%%%%%%%%%%%%%%%%%%%%%%%%%%%%%%%%%%%%%%%%%%%%%%%%%%%%%%%%%%%%%%%%%%%%%%%%%%%%%%%%%%%%%%%%%%%%%%%%%%%%%%%%%%%%%%%%%%%%%%%%%%%%%%%%%%%%%%%%%%%%%%%%%%%%%%%%%%%%%%%%%%%%%%%%%%%%%%%%%%%%%%%%%%%%%%%%%%%%%%%%%%%%%%%%%%%%%%%%%%%%%%%%%%%%%%%%%%%%%%%%%%%%%%%%%%%%%%%%%%%%%%%%%%%%%%%%%%%%%%%%%%%%%%%%%%%%%%%%%%%%%%%%%%%%%%%%%%

\subsubsection{Model II}

As mentioned in Sec. \ref{Shadow}, the critical impact parameter is $b_{c} = 3 \sqrt{3} M$, in model II. Additionally, \(\tilde{a}\) and \(\tilde{b}\) can be formulated as follows
\ie
\Tilde{a} = 3 \sqrt{3} \sqrt{\frac{M^2}{27 M^2+\xi ^2}},
\fe
so that
\ie
\begin{split}
\Tilde{b} = \ln \left(6\right)
+ I_{R}(r_{m}) - \pi.
\end{split}
\fe

Similarly to model I, it is important to highlight that, unlike the Schwarzschild case, the contribution to the parameter \(\tilde{a}\) primarily arises from the effective quantum gravity encoded via $\xi$. Furthermore, \(I_{R}(r_{m})\) can be determined as
\ie
\begin{split}
  I_{R}(r_{m}) &  =   \int_{0}^{1} \mathrm{d}z \left\{ \xi ^2 \left(\frac{1}{27 M \sqrt{M^2 z^2}}-\frac{(z-1)^2 (2 z+1)}{9 \sqrt{3} M \sqrt{-M^2 z^2 (2 z-3)}}\right) \right. \\
  & \left. +\left(\frac{2 \sqrt{3} M}{\sqrt{-M^2 z^2 (2 z-3)}}-\frac{2 M}{\sqrt{M^2 z^2}}\right)  \right\} \\
 & = \frac{540 M^2 \ln \left(3-\sqrt{3}\right)+\xi ^2 \left(-2 \sqrt{3}+9-10 \ln (6)+10 \ln \left(\sqrt{3}+3\right)\right)}{135 M^2}.
\end{split}
\fe
Once again, an analytical result was achieved by considering a small value of \(\xi\). Thus, the deflection angle presented in Eq. (\ref{1d2e3f4l5ections}) is expressed as
\ie
\begin{split}
a(b) = &  -  3 \sqrt{3} \sqrt{\frac{M^2}{27 M^2+\xi ^2}} \ln \left[  \frac{b}{3 \sqrt{3} M} -1     \right] +  \ln \left(6\right) \\
& + \frac{540 M^2 \ln \left(3-\sqrt{3}\right)+\xi ^2 \left(-2 \sqrt{3}+9-10 \ln (6)+10 \ln \left(\sqrt{3}+3\right)\right)}{135 M^2} - \pi \\
& + \mathcal{O} \left\{ \left( b - 3 \sqrt{3} M  \right) \ln\left[\left( b - 3 \sqrt{3} M  \right) \right]  \right\}.
\end{split}
\fe

\begin{figure}
    \centering
     \includegraphics[scale=0.41]{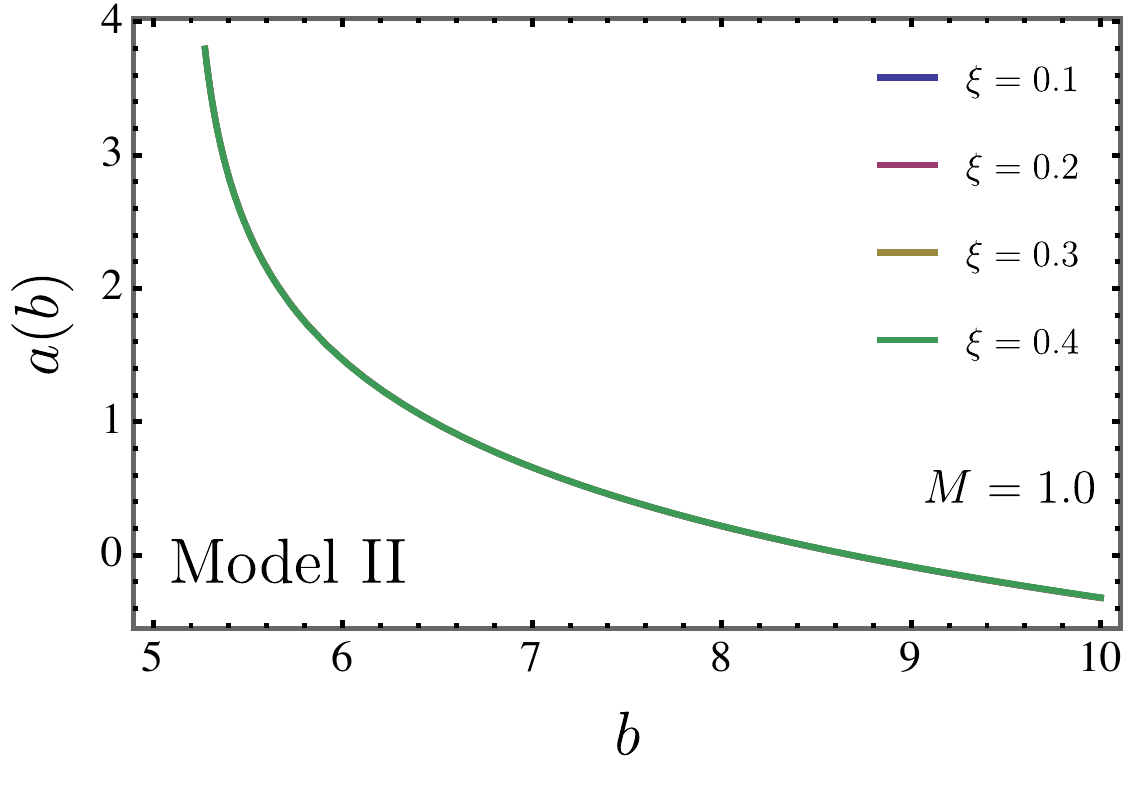}
     \includegraphics[scale=0.45]{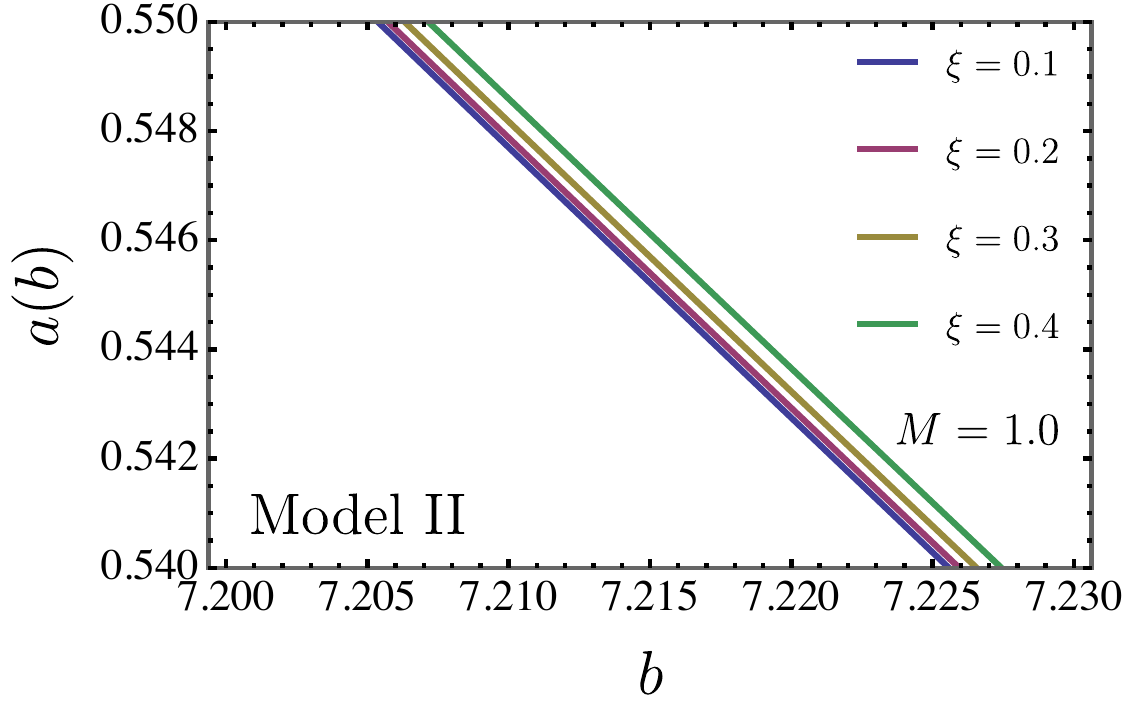}
    \caption{The strong deflection angle (Model II) as a function of $b$ for different values of $\xi$.\\
    %and $M$.
    }
    \label{strongmodelii}
\end{figure}

Fig. \ref{strongmodelii} depicts how the strong deflection angle varies for model II with \(M = 1\) at different values of the parameter \(\xi\). In the left panel, the influence of \(\xi\) is minimal, which aligns with its function as a small quantum correction. Conversely, in the right panel, as \(\xi\) increases, there is a significant increase in the deflection angle. The bottom panel illustrates the deflection angle as a function of \(M\) for a fixed \(\xi = 0.1\), indicating that a larger mass \(M\) leads to a proportionately greater deflection angle.

%%%%%%%%%%%%%%%%%%%%%%%%%%%%%%%%%%%%%%%%%%%%%%%%%%%%%%%%%%%%%%%%%%%%%%%%%%%%%%%%%%%%%%%%%%%%%%%%%%%%%%%%%%%%%%%%%%%%%%%%%%%%%%%%%%%%%%%%%%%%%%%%%%%%%%%%%%%%%%%%%%%%%%%%%%%%%%%%%%%%%%%%%%%%%%%%%%%%%%%%%%%%%%%%%%%%%%%%%%%%%%%%%%%%%%%%%%%%%%%%%%%%%%%%%%%%%%%%%%%%%%%%%%%%%%%%%%%%%%%%%%%%%%%%%%%%%%%%%%%%%%%%%%%%%%%%%%%%%%%%%%%%%%%%%%%%%%%%%%%%%%%%%%%%%%%%%%%%%%%%%%%%%%%%%%%%%%%%%%%%%%%%%%%%%%%%%%%%%%%%%%%%%%%%%%%%%%%%%%%%%%%%%%%%%%%%%%%%%%%%%%%%%%%%%%%%%%%%%%%%%%%%%%%%%%%%%%%%%%%%%%%%

\section{Lenses and observables}

\begin{figure}
    \centering
     \includegraphics[scale=0.5]{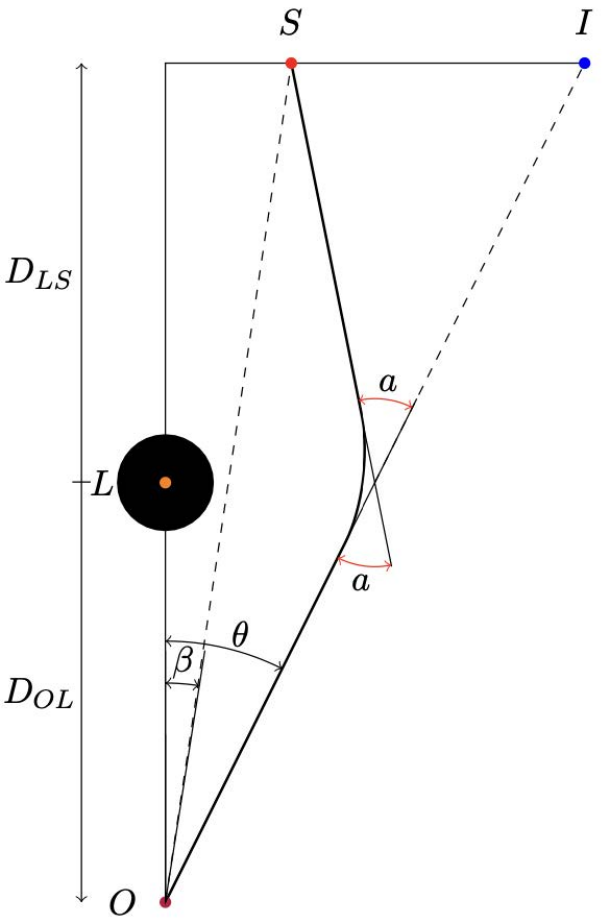}
    \caption{Depiction of gravitational lensing: light emitted from the source \( S \) (represented by the red point) is deflected as it travels towards the observer \( O \) (indicated by the purple point) due to the gravitational influence of a compact object located at \( L \) (shown as the orange point). The observer \( O \) perceives the resulting image at \( I \) (marked in blue). The distance between the lens \( L \) and the observer \( O \) is denoted as \( D_{OL} \), while \( D_{LS} \) represents the distance from the source's projected position to the lens along the optical axis. The large black dot symbolizes the black hole being analyzed \cite{kumar2020gravitational} }
    \label{iluslensing}
\end{figure}

In this section, we will explore several parameters associated with the bending of light within strong gravitational fields, concentrating on our two models. Fig. \ref{iluslensing} serves to illustrate the gravitational lensing phenomenon generated by the black holes outlined in this manuscript. The light emitted from the source, denoted as \(S\) (depicted as a red point), experiences deflection as it moves toward the observer, labeled \(O\) (shown as a purple point), due to the gravitational influence of the black holes located at point \(L\) (represented by an orange point). Moreover, point \(I\) (highlighted in blue) indicates the image seen by observer \(O\). The angular positions of both the source and the observed image are designated as \(\beta\) and \(\theta\), respectively. The angular deviation, represented by \(a\), measures the alteration in the light's trajectory as it passes through the gravitational field.

Additionally, we utilize the same configuration outlined in \cite{bozza2001strong}, where we consider that the source (\(S\)) is nearly perfectly aligned with the lens (\(L\)). This particular setup is significant due to the emergence of relativistic images. In this context, the lens equation that describes the relationship between \(\theta\) and \(\beta\) can be formulated as
\ie
\label{EqLente}
\beta=\theta-\frac{D_{LS}}{D_{OS}}\Delta a_{n}.
\fe
In this context, \(\Delta a_{n}\) denotes the deflection angle that incorporates all the loops made by the photons before arriving at the observer, specifically defined by \(\Delta a_{n}=a-2n\pi\). Within this approach, we utilize the approximation for the impact parameter as follows \(\tilde{b} \simeq \theta D_{OL}\). As a result, the angular deviation can be expressed as
\ie
\label{defle}
a(\theta)=- \Tilde{a}\ln\left(\frac{\theta D_{OL}}{b_c}-1\right)+\Tilde{b}.
\fe

To obtain \(\Delta a_{n}\), we perform an expansion of \(a(\theta)\) around \(\theta = \theta^{0}_n\), where the condition \(\alpha(\theta^{0}_n) = 2n\pi\) is satisfied
\ie
\label{da}
\Delta a_{n}=\frac{\partial a}{\partial\theta}\Bigg|_{\theta=\theta^0_n}(\theta-\theta^0_n) \ .
\fe
Considering Eq. (\ref{defle}) at \(\theta = \theta^{0}_n\), we derive
\ie
\label{To}
\theta^0_{n}=\frac{b_c}{D_{OL}}\left(1+e_n\right), \qquad\text{where}\quad e_n=e^{\Tilde{b}-2n\pi} \ .
\fe

By inserting (\ref{To}) into (\ref{da}), we derive \(\Delta a_{n} = -\frac{\tilde{a} D_{OL}}{b_c e_n} (\theta - \theta^{0}_n)\). When we subsequently integrate this result into the lens equation (\ref{EqLente}), we can formulate the expression for the \(n^{th}\) angular position of the image
\ie
\theta_n\simeq\theta^0_n+\frac{b_ce_n}{\Tilde{a}}\frac{D_{OS}}{D_{OL}D_{LS}}(\beta-\theta^0_n) \ .
\fe

Although the deflection of light preserves surface brightness, the presence of the gravitational lens alters the solid angle of the source, affecting its visual representation. The total flux received from a relativistic image is related to the magnification \(\mu_{n}\), which is defined as \(\mu_n = \left| \frac{\beta}{\theta} \frac{\partial\beta}{\partial\theta} \bigg|_{\theta^0_{n}} \right|^{-1}\). By applying (\ref{EqLente}) and noting that \(\Delta a_{n} = -\frac{\bar{a} D_{OL}}{b_c e_n} (\theta - \theta^{0}_n)\), we derive
\ie
\mu_{n}=\frac{e_n(1+e_n)}{\Tilde{a}\beta}\frac{D_{OS}}{D_{LS}}\left(\frac{b_c}{D_{OL}}\right)^2 \ .
\fe

It is important to note that the magnification factor \(\mu_n\) rises with increasing \(n\). As a result, the brightness from the initial image \(\theta_1\) greatly exceeds that of the later images. However, the total luminosity remains relatively low, largely because of the term \(\left(\frac{b_c}{D_{OL}}\right)^2\). Additionally, a significant aspect is the occurrence of magnification divergence as \(\beta\) approaches zero, highlighting that optimal alignment between the source and the lens enhances the likelihood of observing relativistic images.

In essence, we have detailed the positions and fluxes of relativistic images based on the expansion coefficients (\(\tilde{a}\), \(\tilde{b}\), and \(b_c\)). By redirecting our attention to the inverse problem, we seek to extract these expansion coefficients from observational data. This endeavor not only enhances our understanding of the characteristics of the object causing the gravitational lensing effect but also allows for meaningful comparisons with predictions made by modified theories of gravity.

Furthermore, the impact parameter can be related to \(\theta_{\infty}\), as outlined in \cite{bozza2001strong}
\ie \label{theta}
b_c=D_{OL}\theta_{\infty} \ ,
\fe
where \(\theta_{\infty}\) refers to the additional relativistic images. We will adopt Bozza's approach, as described in \cite{bozza2001strong}, which treats only the outermost image \(\theta_{1}\) as a separate entity, while the other images are included within \(\theta_{\infty}\). To provide further clarification, Bozza proposed the following observables:
  \begin{eqnarray}
       s&=&\theta_{1}-\theta_{\infty}= \theta_{\infty} e^{\frac{\Tilde{b}-2\pi}{\bar{a}}} \ ,\\
       \tilde{r}&=& \frac{\mu_{1}}{\sum\limits_{n=2}^{\infty} \mu_{n} }= e^{\frac{2\pi}{\Tilde{a}}} \ .
       \end{eqnarray}
In these expressions, \( s \) indicates the angular separation, while \(\tilde{r}\) defines the ratio of the flux from the primary image to the total flux of all remaining images. These relationships can be inverted to extract the expansion coefficients. In the following subsection, we will analyze a concrete astrophysical scenario to compute these observables and assess the impact of the effective quantum gravity parameter, \(\xi\), on these values.

%%%%%%%%%%%%%%%%%%%%%%%%%%%%%%%%%%%%%%%%%%%%%%%%%%%%%%%%%%%%%%%%%%%%%%%%%%%%%%%%%%%%%%%%%%%%%%%%%%%%%%%%%%%%%%%%%%%%%%%%%%%%%%%%%%%%%%%%%%%%%%%%%%%%%%%%%%%%%%%%%%%%%%%%%%%%%%%%%%%%%%%%%%%%%%%%%%%%%%%%%%%%%%%%%%%%%%%%%%%%%%%%%%%%%%%%%%%%%%%%%%%%%%%%%%%%%%%%%%%%%%%%%%%%%%%%%%%%%%%%%%%%%%%%%%%%%%%%%%%%%%%%%%%%%%%%%%%%%%%%%%%%%%%%%%%%%%%%%%%%%%%%%%%%%%%%%%%%%%%%%%%%%%%%%%%%%%%%%%%%%%%%%%%%%%%%%%%%%%%%%%%%%%%%%%%%%%%%%%%%%%%%%%%%%%%%%%%%%%%%%%%%%%%%%%%%%%%%%%%%%%%%%%%%%%%%%%%%%%%%%%%%%%%%%%%%%%%%%%%%%%%%

\subsection{Galactic phenomena: gravitational lensing by Sagittarius A$^*$}

Observational data on stellar dynamics provide strong evidence for a compact, enigmatic object residing at the center of our galaxy. This object, identified as the supermassive black hole Sagittarius (Sgr) A\(^*\), is estimated to have a mass of around \(4.4 \times 10^6 M_{\odot}\) \cite{genzel2010galactic}. To gain further insight into this astronomical entity, we investigate its characteristics using the parameter \(\xi\), which allows us to study the behavior of key observables associated with this phenomenon. To compute these observables, we assume a distance of \(D_{OL} = 8.5 \, \text{Kpc}\). Using the critical impact parameters \(b_{c} =  27 \sqrt{\frac{M^4}{27 M^2+\xi ^2}}\) and \(b_{c} = 3 \sqrt{3}M\) for model I and model II respectively, we arrive at the following form for $\theta_{\infty}$ based on Eq. \ref{theta}

\begin{figure}
    \centering    \includegraphics[width=84mm
    ]{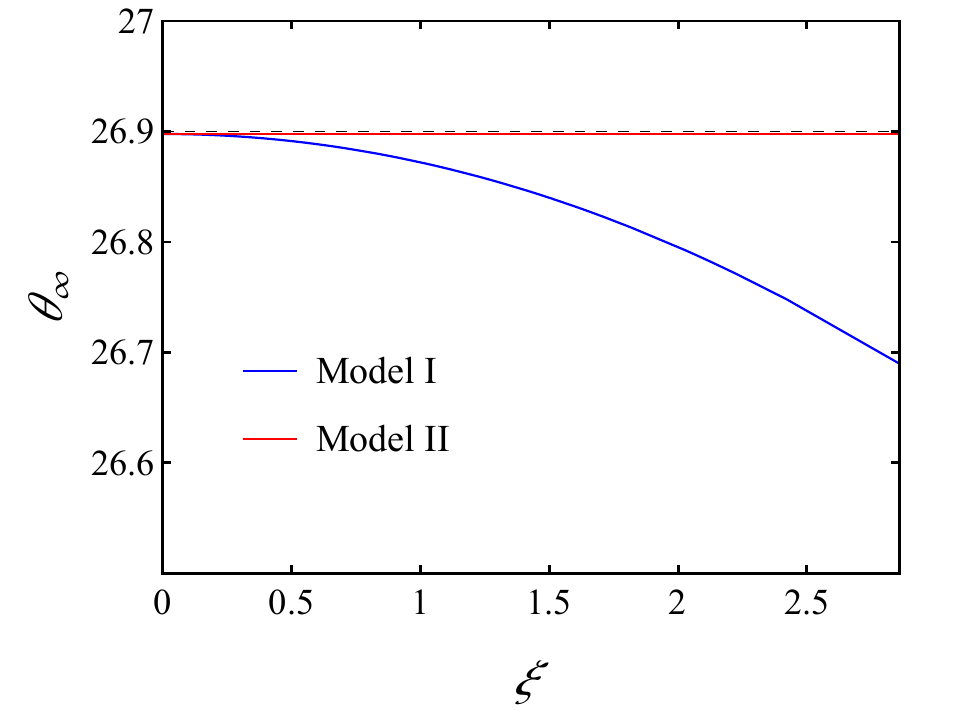}
    \includegraphics[width=80mm]{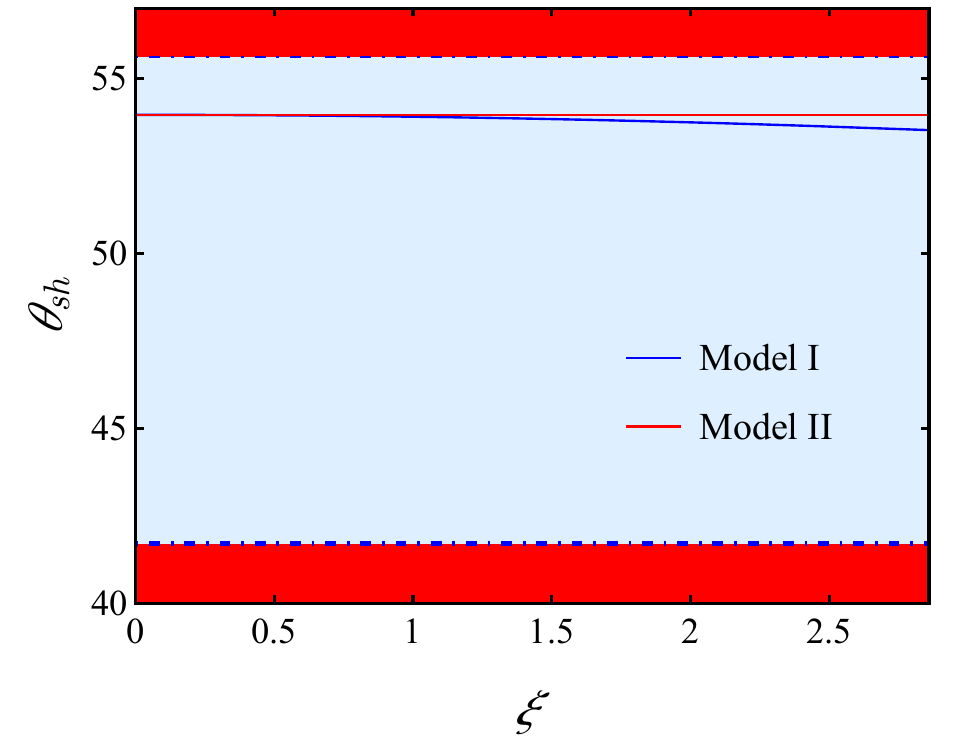}
    \caption{The left panel represents the behavior of lensing observables ($\theta_\infty$) by changing with quantum effective potential parameter $\xi$ for Sagittarius A$^*$. The black dashed corresponds to Sagittarius A$^*$ as a Schwarzschild case. The right panel shows the shadow angular diameter $\theta_{\rm{sh}}$as a function of $\xi$, considering the constraints observed by Sgr A $^*$.}
    \label{fig:angel}
\end{figure}

\ie
\begin{split}
  \theta^I_{\infty} = 614.965 \sqrt{\frac{1}{\xi ^2+522.72}} \approx &26.8977-0.0257286 \xi ^2+O\left(\xi ^4\right),\\
  \theta^{II}_{\infty} \approx & 26.8977.
\end{split}
\fe
In Fig. \ref{fig:angel}, the behaviour of observable, $\theta_\infty$ with variation of $\xi$, is shown in the left panel. It can be observed that in model I, the angular position, slightly deviates from the result obtained for the Schwarzschild black hole. However, it matches the value obtained for the Schwarzschild black hole in model II due to the point that the impact parameter has the same value as the Schwarzschild case. Moreover the shadow angular diameters $\theta_{\rm{sh}}=2\theta_{\infty}$ are represented in the right panel. Also the $1\sigma$ contaraints of Sqr. A$^*$ determined the allowed range of angular shadow diameter as $41.7 < \theta_{\rm{sh}} < 55.6$ ($\mu as$) according to ref. \cite{akiyama2022first}. Therefore there isn't any constraint on the effective quantum gravity parameter based on the right panel plots, and angular shadow diameter lies in the allowed range for both models.

%%%%%%%%%%%%%%%%%%%%%%%%%%%%%%%%%%%%%%%%%%%%%%%%%%%%%%%%%%%%%%%%%%%%%%%%%%%%%%%%%%%%%%%%%%%%%%%%%%%%%%%%%%%%%%%%%%%%%%%%%%%%%%%%%%%%%%%%%%%%%%%%%%%%%%%%%%%%%%%%%%%%%%%%%%%%%%%%%%%%%%%%%%%%%%%%%%%%%%%%%%%%%%%%%%%%%%%%%%%%%%%%%%%%%%%%%%%%%%%%%%%%%%%%%%%%%%%%%%%%%%%%%%%%%%%%%%%%%%%%%%%%%%%%%%%%%%%%%%%%%%%%%%%%%%%%%%%%%%%%%%%%%%%%%%%%%%%%%%%%%%%%%%%%%%%%%%%%%%%%%%%%%%%%%%%%%%%%%%%%%%%%%%%%%%%%%%%%%%%%%%%%%%%%%%%%%%%%%%%%%%%%%%%%%%%%%%%%%%%%%%%%%%%%%%%%%%%%

\section{Conclusion}
\label{conclusion}

In this study, we investigated the implications of black holes within the framework of effective quantum gravity as proposed by Ref. \cite{zhang2024black}. First, we analyzed the behavior of the lapse function and the structure of the horizons. The trajectories of light in the spacetimes of Models I and II were numerically simulated for various values of \(\xi\). Additionally, we examined wave-like phenomena using a test scalar field and computed the absorption cross-section through partial wave analysis for both models. Our findings revealed that, in both cases, the effective quantum gravity parameter \(\xi\) impacted the partial and total absorption cross--sections, with higher values of \(\xi\) leading to lower absorption as observed by a distant detector. However, model II showed a reduced sensitivity to variations in \(\xi\). We also demonstrated that at low frequencies, the total absorption cross--section approached the black hole’s horizon area, while at higher frequencies, it converged with the capture cross-section for null geodesics, as predicted by partial wave analysis.

We further explored the greybody factor bounds for scalar and Dirac perturbations. For scalar fields in both models, we found that the greybody factor decreased as \(\xi\) increased, suggesting that stronger effective quantum gravity reduces the probability of wave penetration through the black hole's potential barrier. In contrast, for the Dirac field, the greybody factor in model I remained unaffected by changes in \(\xi\), whereas in model II, it tended to decrease as \(\xi\) increased.

Finally, we carried out a comprehensive analysis of gravitational lensing under both weak and strong deflection limits. For the weak deflection regime, the \textit{Gauss--Bonnet} theorem was applied, whereas the \textit{Tsukamoto} method was utilized for the strong deflection limit. Additionally, we derived the corresponding observables and presented an astrophysical application using data from the Event Horizon Telescope for Sgr A*. In summary, for the model I, \(\theta_{\infty}\) displayed a slight deviation from the value found for the Schwarzschild black hole. In contrast, for model II, the effective quantum corrections encapsulated by the parameter \(\xi\) did not influence \(\theta_{\infty}\).

As a future direction, exploring particle creation for fermions and bosons, accretion disk of matter, and applying the corrected Newman--Janis technique to obtain an axisymmetric solution are promising topics for further research. These and other related ideas are currently under investigation.

%%%%%%%%%%%%%%%%%%%%%%%%%%%%%%%%%%%%%%%%%%%%%%%%%%%%%%%%%%%%%%%%%%%%%%%%%%%%%%%%%%%%%%%%%%%%%%%%%%%%%%%%%%%%%%%%%%%%%%%%%%%%%%%%%%%%%%%%%%%%%%%%%%%%%%%%%%%%%%%%%%%%%%%%%%%%%%%%%%%%%%%%%%%%%%%%%%%%%%%%%%%%%%%%%%%%%%%%%%%%%%%%%%%%%%%%%%%%%%%%%%%%%%%%%%%%%%%%%%%%%%%%%%%%%%%%%%%%%%%%%%%%%%%%%%%%%%%%%%%%%%%%

\section*{Acknowledgements}
A. A. Araújo Filho is supported by Conselho Nacional de Desenvolvimento Cient\'{\i}fico e Tecnol\'{o}gico (CNPq) and Fundação de Apoio à Pesquisa do Estado da Paraíba (FAPESQ) - [150891/2023-7]. A. {\"O}. and R. P. would like to acknowledge networking support of the COST Action CA18108 - Quantum gravity phenomenology in the multi-messenger approach (QG-MM), COST Action CA21106 - COSMIC WISPers in the Dark Universe: Theory, astrophysics and experiments (CosmicWISPers), the COST Action CA22113 - Fundamental challenges in theoretical physics (THEORY-CHALLENGES), and the COST Action CA21136 - Addressing observational tensions in cosmology with systematics and fundamental physics (CosmoVerse). A. {\"O}. would also thank TUBITAK and SCOAP3 for their support.

\bibliography{main}
\bibliographystyle{unsrt}

\end{document}